\documentclass[fleqn,10pt]{wlscirep}
\usepackage[utf8]{inputenc}
\usepackage[T1]{fontenc}
\usepackage{mathrsfs}
\usepackage{tabularx}
\usepackage{booktabs}
\title{Superconducting Phases in Lithium Decorated Graphene LiC$_6$}

\author[1]{Rouhollah Gholami}
\author[1,2,*]{Rostam Moradian}
\author[3]{Sina Moradian}
\author[4]{Warren E. Pickett}
\affil[1]{Physics Department, Faculty of Science Razi University, Kermanshah, Iran}
\affil[2]{Nano science and nano technology research center,  Razi University, Kermanshah, Iran}

\affil[*]{Correspondence and requested materials should be addressed to R. M. [rmoradian@raz.ac.ir]}

\affil[3]{Department of Electrical and Computer Engineering, University of Central Florida, Orlando, Florida, USA.}
\affil[4]{Department of Physics UC Davis, One Shield Avenue, Davis, CA 95616, USA.} 

\begin{abstract}
A study of possible superconducting phases of graphene has been constructed in
detail. A realistic tight binding model, fit to ab initio calculations, accounts for the Li-decoration
of graphene with broken lattice symmetry, and includes $s$ and $d$ symmetry
Bloch character that influences the gap symmetries that can arise. 
The resulting seven hybridized Li-C orbitals that support nine possible bond pairing
amplitudes.  The gap equation is solved for all possible gap symmetries. One band
is weakly dispersive near the Fermi energy along $\Gamma\rightarrow M$ where its Bloch wave
function 
 has linear combination of $d_{x^2-y^2}$ and $d_{xy}$ character, and is responsible for
$ d_{x^{2}-y^{2}}$ and $d_{xy}$ pairing with lowest pairing energy in our model.
These symmetries almost preserve properties from a two band model of pristine graphene.  
Another part of this band, along $K\rightarrow \Gamma$, is nearly degenerate with upper $s$ band that favors
extended $s$ wave pairing which is not found in two band model.
Upon electron doping to a critical chemical potential
$\mu_1=0.22 eV$ the pairing potential decreases, then increases until a second critical value
$\mu_2$=1.3 eV at which a phase transition to a distorted $s$-wave occurs. The distortion of  $d$- or s-wave phases are a consequence of decoration  which is not appear in two band pristine model. 
In the  pristine graphene  these phases   convert to usual d-wave or extended $s$-wave pairing. 
\end{abstract}
\begin{document}

\flushbottom
\maketitle
%
%
\thispagestyle{empty}
\section{Introduction}
Two dimensional superconducting phases have become of great interest 
since the discovery of the high temperature superconducting (HTS) cuprates 
and subsequent finding of Fe-pnictide and -chalcogenide HTSs.
Interest was re-invigorated by the discovery of superconductivity onsets
up to 75K in single layer FeSe grown on $SrTiO_3$ and related
substrates.\cite{Wang2012, He2013} With the enormous research activity focused on
graphene in recent years, it is not surprising that graphene-based superconductivity
has become an active area of research. Very recently superconductivity up to 1.7K
has been reported\cite{Cao2018} in magic angle bilayer graphene, which will
buttress activity on two dimension superconductors and especially the related
type that we discuss here.

Superconductivity has been known for some time in intercalated graphite compounds such 
as $C_{6}Ca$ and $C_{6}Yb$ \cite{Weller-naturephysics2005}. With the many remarkable
properties of graphene, it has been anticipated that doping by gating or by decorating with
electro-positive elements, thereby moving the chemical potential away from the Dirac points, 
might induce superconductivity.  However, graphene decorated with alkali metals has three 
valence bands with  one weakly dispersive band near Fermi energy. Due to this flat band, 
there are additional available 
states around the Fermi level and the required pairing potential is reduced. 

Discussion of superconductivity in doped graphene has been primarily within theoretical
models, as we review below, but some encouraging data have been reported. 
Experimental evidence for a superconducting gap in
Li-decorated monolayer graphene around 6 K has been reported by Ludbrook 
{\it et al.} based on angle-resolved photoemission spectroscopy\cite{Ludbrook2015} (ARPES).
Scanning tunneling spectroscopy (STS) was applied by Palinkas {\it et al.}\cite{Palinkas2017}
to graphene suspended on tin nanoparticles, who concluded that superconductivity is
induced in the graphene layer. Evidence of superconductivity in Li-decorated 
few layer graphene at 7.4 K
has been reported by Tiwari and collaborators\cite{Tiwari2017}.  
Low temperature mobility of K and Li atoms on graphene was observed by Woo {\it et al.},
and suggest that mobility may persist at lower temperatures,\cite{Woo2017} 
which would provide new challenges for theory.

Various mechanisms of pairing have been proposed. Uchoa and Castro-Neto modeled pristine and
doped graphene with electron-phonon coupling or plasmon mediated  in mind\cite{Uchoa}. Repulsive electron-electron 
interactions were modeled by Nandkishore and collaborators.\cite{Nandkishore, Nandkishore2014} 
Beginning from pristine 
graphene, varying the chemical potential leads to dominant chiral singlet $d_{x^2 -y^2}+i d_{xy}$
pairing for nearest neighbor interaction, according to Black-Schaffer 
{\it et al.}\cite{Blackschaffer2007}  a triplet $f$-wave state has been proposed to arise from
next-nearest neighbor interaction with chemical potential near van Hove peak.\cite{Kiesel} 
Both chiral and conventional $p$-wave states in graphene have been discussed\cite{Ma}, 
with the many pictures raising various possibilities but little of a certain nature. 

More specific predictions have begun to appear.
Profeta {\it et al.} predicted\cite{Profeta} based on Eliashberg theory that decoration
by electron donating atoms such as Ca and Li would make single layer graphene
superconducting,  with modest critical
temperatures in the 1-8 K range. In somewhat related work, 
Wong {\it et al.} have predicted\cite{Wong2017}
from an ab initio treatment a critical temperature around T$_c$=14K 
for carbon nanotubes, which was increased to above
100 K for a certain type of carbon ring.

Expectations of adjusting the chemical potential include gating, but the main focus has
been on decoration of graphene by electropositive atoms, viz. alkalis or alkaline earths.
Charge migration from such decorating atoms to the graphene layer will affect the
C-C bonding, leading to contraction or expansion of the graphene hexagons that are 
centered by the decorating atoms, thus breaking the symmetry of C-C hopping integrals 
around the honeycomb loop.  This asymmetric graphene
layer will be referred to in this paper as ``shrunken graphene''.
Taking LiC$_6$ for illustration, each cell site has six C atoms in a hexagon with an alkaline atom
lying above the center of the hexagon. The C $\pi$ orbitals and alkaline atom's $s$-orbital 
hybridize to give seven ``molecular'' orbitals. 
For two dimensional graphene-like structures effects, differences in nearest neighbour 
hopping integrals affect the  band structure near the important Dirac point, 
which is folded back to the $\Gamma$ point of the shrunken graphene 
superlattice investigated by Hou {\it et al.}\cite{Hou} and Long-Hua 
{\it et al.}\cite{ Long-Hua}. For such systems not even the full analytic 
tight-binding band structure has yet been reported. The intent here is to extend
study of this system, with representative LiC$_6$, from the underlying electronic structure
to investigation of the possible superconducting phases.

The organization of the paper is as follows.
In Sec. II the interacting seven orbital model Hamiltonian is presented. 
The exact band structure of the normal state of this shrunken graphene
system is described in Sec. III.  Perturbation theory is applied to obtain the
band structures in analytic form. Applying the Hubbard model and minimizing free 
energy of the superconductor state, we obtain in Sec. IV the gap equations and 
approximate critical temperature. These equations are solved analytically to 
establish the possible pairing symmetries and other properties of the superconducting
states. A summary is provided in Sec. V.
\section{Model Hamiltonian}
\begin{figure}[ht]
\centering
\includegraphics[width=14cm]{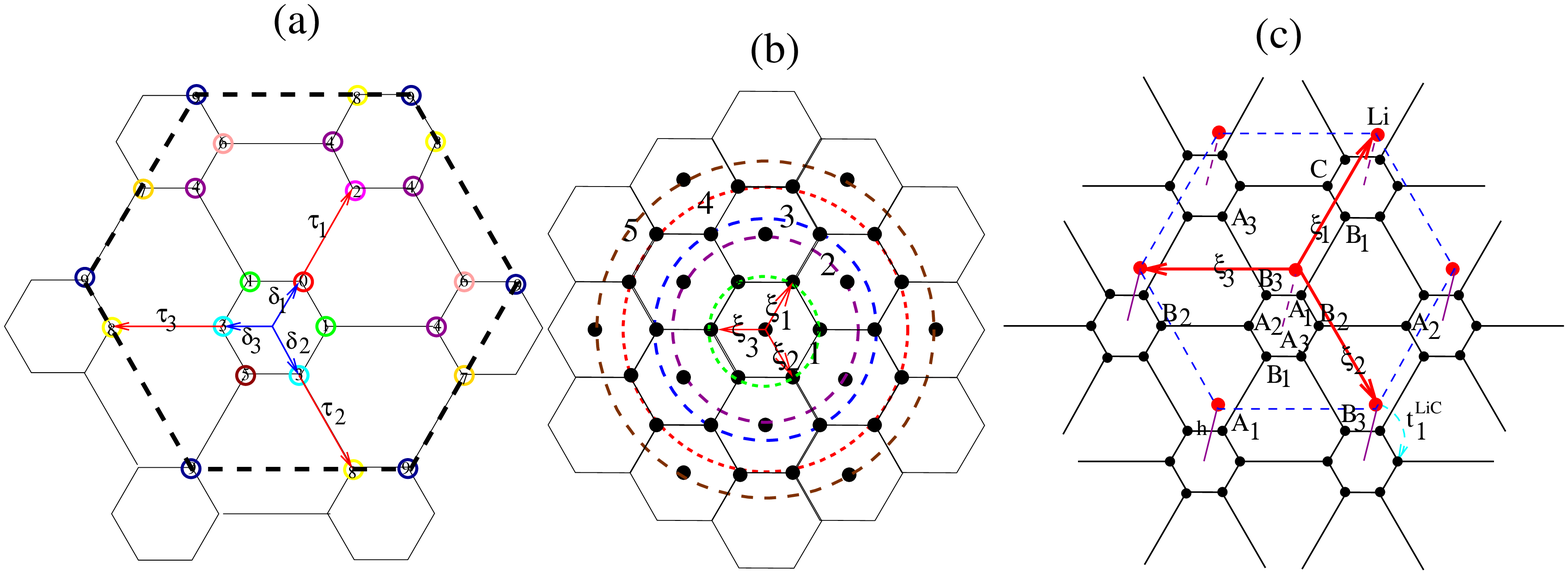}
\caption{(Color online) (a) Schematic diagram of the shrunk graphene lattice,
with the distortion emphasized. (b) The hexagonal Li sheet, indicating the
circles that Li neighbors lie on. 
(c) Diagram of the graphene decorated by lithium. The red Li atoms lie above
the centers of the C hexagons.  }
 \label{figure:graphene decorated by lithium} 
\end{figure}

Because the unit cell contains several atoms with important specific aspects,
we provide many of the details of the expressions that can be obtained analytically.
LiC$_6$, as illustrated in Fig.~\ref{figure:graphene decorated by lithium}, consists 
of a graphene layer decorated  by a lithium layer in which Li atoms are located 
at the center of  a carbon hexagon surrounded by six empty center hexagons. 
The height of Li above the carbon layer
is calculated to be $h_{z}= 1.85\AA$, somewhat smaller than the value 1.93\AA~ obtained by
Profeta {\it et al.}\cite{Profeta}. The nearest Li-C distances are $h=2.40\AA$.
Since the Li $2s$ orbital energy is higher than the C $2p_{z}$ orbital, charge 
transfer occurs. It is calculated that $0.685 e$ from Li transfers to the six C atoms
equally.\cite{Guzman}  The positive Li ion and negative C ion provide a relative
Coulomb (Madelung) shift in site potentials of the two atoms.

The attractive interaction  between Li and C ions after charge transfer 
contracts the Li-C distance and
reduces the C-C bond lengths in the Li-centered hexagon to $a_{1}=1.425\AA$, while the bond 
length of nearest neighbor C atoms in different hexagons is slightly larger at
$a_{2}=1.426\AA$.  For Ca instead of Li, this difference should be larger, hence we
keep these lengths distinct. The hopping integral between short-bond carbons is $t_{1}$, 
with that between stretched carbon sites is denoted $t^{'}_{1}$.  We refer to  this 
broken symmetry situation as ``shrunk graphene''. The difference in hopping
amplitudes indicates that the new Li-C hopping parameter is the central new
feature in LiC$_6$ compared to graphene.  Symmetry breakdown leads to
the opening of a small energy gap at the $\Gamma$ point.

The lattice then becomes a two dimensional hexagonal Bravais lattice with seven 
atomic sites. These will be labeled as $A_1, A_2, A_3, B_1, B_2, B_3$ and $Li$,
as illustrated in Fig.~\ref{figure:graphene decorated by lithium}. 
The Hamiltonian of this system is 
\begin{eqnarray}
 \hat{H}=-\sum_{i\alpha}\sum_{j \beta,\sigma } 
    t_{i\alpha,j\beta}^{\sigma,\sigma}{\hat{c}}_{i\alpha\sigma}^{\dagger}{\hat{c}}_{j\beta\sigma} 
    +\sum_{i\alpha ,\sigma }(\epsilon_{i\alpha}-\mu_{o} ) \hat{n}_{i\alpha\sigma} 
    +\frac{1}{2}\sum_{i\alpha,\sigma}\sum_{j \beta ,\sigma{}'}U _{i\alpha,j \beta}^{\sigma ,\sigma{}'} 
      \hat{n}_{i\alpha\sigma}   \hat{n}_{j\beta\sigma{}'} = {\hat H}_{N} + {\hat H}_{P}.
\label{eq:full-Hamiltonian}
\end{eqnarray}
Here $H_N$ and $H_P$ denote the non-interacting and interaction Hamiltonians
respectively. In these expressions $\alpha$ and $\beta$ run over $ A_i$, $B_i$ and $Li$. Here 
${\hat{c}}_{i\alpha\sigma}^{\dagger}$, ${\hat{c}}_{i\alpha\sigma} $ are 
creation and annihilation operators of an electron with spin $\sigma$ on 
subsite $\alpha$ of $i$th lattice site, and
$\hat{n}_{i\alpha\sigma}={\hat{c}}^{\dagger}_{i\alpha\sigma}{\hat{c}}_{i\alpha\sigma} $ 
is the electron number operator.
The noninteracting chemical potential is $\mu_{0}$
and $t_{i\alpha,j \beta}$ is the hopping integral from the $\alpha$ site 
of $i$th cell to the $\beta$ site of $j$th cell.
We denote the on-site energy by $\epsilon_{\alpha}$.

The interaction stated above corresponds to an extended (negative $U$) Hubbard
model, which allows a variety of phenomenological values to be chosen
and studied. It is largely for this reason that we provide substantial detail
of the underlying, non-interacting C-Li lattice and electronic structure. The
interactions that we study are introduced in Sec. IV. 
\section{Normal state of LiC$_6$}
Many studies of graphene rely on tight binding parametrization of the band structure.
The early parametrization of Wallace\cite{Wallace1947} already employed both
first and second neighbors. Extensions in various ways have
followed,\cite{Reich2002,Kundu2011} culminating in the application of Wannier
functions by Jung and MacDonald\cite{Jung2013} to provide simple but realistic
five parameter model and a more accurate but more involved 15 parameter model.  Our aim in this section is to construct a realistic seven band model for distorted LiC$_6$, while also developing the formalism to allow exploration of superconducting
phases once the interaction has been included.

The distortion of the graphene layer to shrunken graphene and the coupling
to Li requires a considerable generalization of the underlying tight binding
model Hamiltonian, and many of the details are
relegated to appendices.
The Hamiltonian of non-interacting LiC$_6$ is 
\begin{equation}
 \hat{H}_{N}=-\sum_{i\alpha}\sum_{j \beta,\sigma } t_{i\alpha,j\beta}^{\sigma,\sigma}c_{i\alpha\sigma}^{\dagger}c_{j\beta\sigma} +\sum_{i\alpha ,\sigma }(\epsilon_{i\alpha}-\mu_{o} ) \hat{n}_{i\alpha\sigma}.
\label{eq:H-normal}
\end{equation}
Eq.~\ref{eq:H-normal} incorporates broken symmetries in the on-site energies, 
hopping integrals, and bond lengths.
Here, it has been assumed  that on site energies $\epsilon_{A_{i}}=\epsilon_{A}$ and $\epsilon_{B_{i}}=\epsilon_{B}$.  It is diagonalized in terms of Bloch eigenfunction of the form 
Eq.~A.2
.  In matrix representation, the equation for the coefficients becomes
  \begin{eqnarray}
 \left( {\begin{array}{*{20}{c}}
{{\varepsilon _0}(\vec k)}&\vline& d_{c1}(\vec k)&d_{c3}(\vec k)&d_{c2}(\vec k)&\vline& d_{c1}^{*}(\vec k)&d_{c3}^{*}(\vec k)&d_{c2}^{*}(\vec k)\\
\hline
d_{c1}^{*}(\vec k)&\vline& {{\varepsilon _1}(\vec k)}&{ \beta (\vec k)}&{ \gamma (\vec k)}&\vline& {{\tau_1}(\vec k)}&{{d_2(\vec k)}}&{{d_3(\vec k)}}\\
d_{c3}^{*}(\vec k)&\vline& {  {\beta ^*}(\vec k)}&{{\varepsilon _1}(\vec k)}&{  \theta (\vec k)}&\vline& {{d_2(\vec k)}}&{{\tau_3}(\vec k)}&{{d_1(\vec k)}}\\
d_{c1}^{*}(\vec k)&\vline& {  {\gamma ^*}(\vec k)}&{  {\theta ^*}(\vec k)}&{{\varepsilon _1}(\vec k)}&\vline& {{d_3(\vec k)}}&{{d_1(\vec k)}}&{{\tau_2}(\vec k)}\\
\hline
d_{c1}(\vec k)&\vline& {\tau_1^*(\vec k)}&{d_2^{*}(\vec k)}&{d_3^*(\vec k)}&\vline& {{\varepsilon _2}(\vec k)}&{  {\beta ^*}(\vec k)}&{  {\gamma ^*}(\vec k)}\\
d_{c3}(\vec k)&\vline& {d_2^{*}(\vec k)}&{\tau_3^*(\vec k)}&{d_1^{*}(\vec k)}&\vline& {  \beta (\vec k)}&{{\varepsilon _2}(\vec k)}&{  {\theta ^*}(\vec k)}\\
d_{c2}(\vec k)&\vline& {d_3^*(\vec k)}&{d_1^{*}(\vec k)}&{\tau_2^*(\vec k)}&\vline& {  \gamma (\vec k)}&{  \theta (\vec k)}&{{\varepsilon _2}(\vec k)}
\end{array}} \right)\left( {\begin{array}{*{20}{c}}
{{\mathscr{C}_{0}}({E_i}(\vec k))}\\
\hline
{{\mathscr{C}_{{1}}}({E_i(\vec k)})}\\
{{\mathscr{C}_{{2}}}({E_i}(\vec k))}\\
{{\mathscr{C}_{{3}}}({E_i}(\vec k))}\\
{{\mathscr{C}_{{4}}}({E_i}(\vec k))}\\
{{\mathscr{C}_{{5}}}({E_i}(\vec k))}\\
{{\mathscr{C}_{{6}}}({E_i}(\vec k))}
\end{array}} \right)={E_i(\vec k)} \left( {\begin{array}{*{20}{c}}
{{\mathscr{C}_{0}}({E_i}(\vec k))}\\
\hline
{{\mathscr{C}_{{1}}}({E_i(\vec k)})}\\
{{\mathscr{C}_{{2}}}({E_i}(\vec k))}\\
{{\mathscr{C}_{{3}}}({E_i}(\vec k))}\\
{{\mathscr{C}_{{4}}}({E_i}(\vec k))}\\
{{\mathscr{C}_{{5}}}({E_i}(\vec k))}\\
{{\mathscr{C}_{{6}}}({E_i}(\vec k))}
\end{array}} \right)
\label{eq:matrix form- full-sch}
\end{eqnarray} 
where  $d_{ci}(\vec k)$, $\varepsilon_{i}(\vec {k})$, $\beta({\vec k})$, $\theta({\vec k})$, $\gamma({\vec k})$,  $d_{i}(\vec k)$ and $\tau_{i}(\vec k)$ functions are defined in supplementary materials
 Eqs.~A.7,~A.8,~A.9, and A.10
  respectively. 
  For general ${\vec k}$ vectors,  it is challenging to obtain an exact analytical expression 
for the full Hamiltonian in Eq.~\ref{eq:matrix form- full-sch} and it would not be transparent anyway.
 However, analytical expression for Eq.~\ref{eq:matrix form- full-sch} can be achieved  in two steps.
Since hopping from Li atoms to nearest neighbor carbon sites $t_{1}^{LiC}$ is  small with respect to C-C nearest neighbor hopping  $t_1$, 
  by  first neglecting the lithium-carbon hopping $t_{1}^{LiC}\rightarrow 0$, first column and row of the Hamiltonian matrix in Eq.~\ref{eq:matrix form- full-sch} are omitted, 
 the remaining part   given by
Eq.~B.1 is uncoupled shrunken graphene Hamiltonian  which can be diagonalized exactly to obtain $E_{sh,n}$. Finally, 
  Li-C coupling is taken into account by perturbation theory
   to obtain eigenvalues $E_{n}$, as presented in the appendices.

 \subsection{Uncoupled $C_6$ Dispersion Relations }
\begin{figure}[ht]
\centering
\includegraphics[width=8cm]{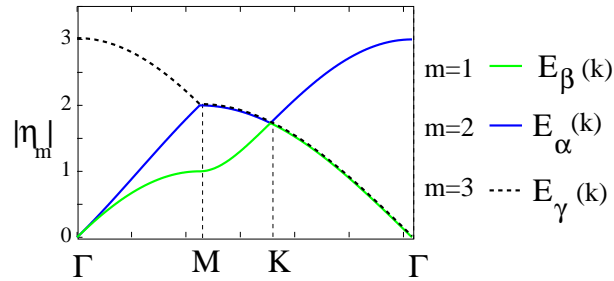}
\caption{(Color online) A plot of the dispersion expressions $|\eta_{m}(\vec k)|$, the 
three folded branches of pristine 
 $\pi^{*}$  band structure in the mini-Brillouin zone of  graphene C$_6$. The Bloch wave character 
 of $E_{\beta}$ is $f_d^1|d-ip>+f_p^1|p-id>$ , of the  $E_{\alpha}$ is $f_d^2|d+ip>+f_p^2|p+id>$ 
 and for $E_{\gamma}$ is $f_s|s>+f_f|f>.$
Here we use abbreviated notation $f_{d}^{1(2)}|
d\pm i p>=f_{d_{x^2-y^2}}^{1(2)}( |d_{x^2-y^2} > \pm i|p_{x}>)$ and $f_{p}^{1(2)}|
p\pm i d>=f_{p_{y}}^{1(2)}( |p_{y} > \pm i|d_{xy}>)$} 
 \label{figure:eta-k} 
\end{figure}

By first  neglecting the lithium-carbon hopping, $t_{1}^{LiC}\rightarrow 0$, the uncoupled shrunken graphene Hamiltonian given by 
  Eq.~B.1
 can be diagonalized exactly. Even though Li-C hopping has been neglected but still remaining part of shrunken Hamiltonian in the most general case, include broken symmetries in the  
hopping integrals, bond lengths and on-site energies.
  The non trivial eigenvalues of uncoupled shrunken graphene Hamiltonian in general form are given by 
  \begin{eqnarray}
{E_{sh,ml}}(t_{i},\vec{\xi_{i}},\vec{k}) =- {\mu _{o}} +\alpha (\vec k)  +u_{m}\Pi_{0}(\vec k)+ u^ *_{m} \Pi^{*}_{0}(\vec k) + \frac{1}{2}\left[ {{\varepsilon _A} + {\varepsilon _B} + {{( - 1)}^l}\sqrt {{{\left( {{\varepsilon _A} - {\varepsilon _B}} \right)}^2} + 4{w_{m}}(\vec k)}   } \right]
\label{eq:band-structure of shrunk graphene-text}
\end{eqnarray}
with details presented in supplementary materials Appendix B. However, the obtained equations are often complicated. To provide insight into the method, uncoupled shrunken graphene Hamiltonian  can diagonalized in some  particular cases. 
The Brillouin zone (BZ) of C$_6$ is one third of that of graphene, with the Dirac points folded back to the $\Gamma$  point. In this mini-BZ, the two $\pi$ bands of pristine graphene i.e. $E_{\pm}=\pm t_1 |\eta_0|$ folds to six branches as illustrated inFig.~\ref{figure:eta-k}. These branches are solutions of  
Eq.~B.1 in the limited case of pristine which  in the nearest neighbor approximation they are given by,
  \begin{eqnarray}
  E_{\gamma}^{\pm}(\vec{k})&=&\pm t_1 |\eta_0(\vec{k})|,~~~~\eta_0(\vec{k})=
  e^{i\vec{k}.\vec{\delta_1}} +
  e^{i\vec{k}.\vec{\delta_2}}+e^{i\vec{k}.\vec{\delta_3}}\nonumber\\
   E_{\beta}^{\pm}(\vec{k})&=&\pm t_1 |\eta_1(\vec{k})|,~~~~\eta_1(\vec{k})=e^{i\vec{k}.\vec{\delta_1}} +
  e^{i\frac{4\pi}{3}}e^{i\vec{k}.\vec{\delta_2}}+e^{i\frac{2\pi}{3}}e^{i\vec{k}.\vec{\delta_3}}\nonumber\\
   E_{\alpha}^{\pm}(\vec{k})&=&\pm t_1 |\eta_2(\vec{k})|,~~~~\eta_2(\vec{k})=e^{i\vec{k}.\vec{\delta_1}} +
 e^{i\frac{2\pi}{3}} e^{i\vec{k}.\vec{\delta_2}}+e^{i\frac{4\pi}{3}}e^{i\vec{k}.\vec{\delta_3}}.
 \label{eq:pristine-band-branches}
\end{eqnarray}   
 Exact analytical solutions  for pristine graphene wherein  next neighbor hopping integrals are taken into account are presented in
supplementary materials Eqs.~B.7 and B.8. As  shown  in Fig.~\ref{figure:eta-k} one sees that $E_{\beta}^{\pm}(\vec{k})$  is weakly dispersive near the van Hove singularity at the  saddle points $M$ at 3/8 or 5/8 filling (0.25 electron per carbon doping), this band plays a major role in the formation of superconductivity in graphene. Also, one can observe that the band structure is four-fold degenerate at  the charge neutral Dirac points.  Solution of the Schr\"odinger equation for pristine graphene in the mini-BZ has another advantage: the Bloch-wave symmetry character of each branch can be distinguished. The Bloch coefficients of the branch  labeled by $E_\gamma$  are of $s$-wave  character, $C_{A_i}=(1,~~1,~~1)$ while for those labeled as $E_\alpha$  and $E_\beta$  are of the form $d\pm id$ -wave i.e. $C_{A_i}=(1,~~e^{\pm i\frac{i2\pi}{3}},~~e^{\pm i\frac{i4\pi}{3}})$ as illustrated inFig.~\ref{figure:eta-k} and
 demonstrated in more detail in  Appendix B ,
 Eqs.~B.4 and B.6. 
 This becomes important when it is shown that different superconducting phases of graphene in a variety of doping regimes are due to electron pairing in each of these branches.  
 
  Decoration of graphene with metals reduces symmetries that lead to removal of bands degeneracy in some regions. While  decoration causes expansion and contraction of  bonds length in three inequivalent directions in the honeycomb lattice  i.e. $|\vec{\tau_i}|\neq|\vec{\delta_i}|$,   
  eigenenergies $E_{sh,ml}(t_{i},\vec{\xi_{i}},\vec{k})$ in Eq.~\ref{eq:band-structure of shrunk graphene-text}  
do not depend on the bond lengths
 $\vec{\tau_i}$ and $\vec{\delta_i}$ separately but are functions of $LiC_6$ lattice bases length
 $|\vec{\xi_i}=\vec{\tau_i}+2\vec{\delta_i}|$,  so symmetry breakdown of bond lengths 
does not break symmetries of bands. Symmetry reduction of 
hopping integrals removes degeneracies occurring in pristine graphene band structure,
with the most important effect being to open a gap $E_{g}=2|t^{'}_{1}-t_{1}|$ at the Dirac 
point which has been folded back to the $\Gamma$ point. This gap arises from symmetry breaking of the nearest neighbor  hopping and dose not affected by the other next neighbors hopping nor by the Li-C hopping integral. Comparison with DFT band structures gives $E_g=0.36eV$. 
Another   gap can arise at the $\Gamma$ point because of symmetry breaking of on-site energies $\varepsilon_A\neq \varepsilon_B$, seen from Eq.~\ref{eq:band-structure of shrunk graphene-text}. 
For the case $t_1=t_1^{'}$  the gap becomes $2|\varepsilon_A-\varepsilon_B|$. 
In Li decorated graphene that we consider here, all carbon on-site energies are equal so this type gap does not arise.

 While for folded but pristine graphene Bloch wave solutions are pure $s$-wave or chiral $d\pm id$-wave
and there are no mixed states, when symmetries in hopping integrals are broken by decoration, Bloch functions are linear combinations of all  these phases,
 Eq.~B.6.
Equation~B.7
demonstrates that for a general $\vec{k}$ all probabilities are equal in pristine graphene i.e.
$|C_{A_i}(E_m)|^2=|C_{B_i}(E_m)|^2=\frac{1}{6}$. In shrunken graphene these 
probabilities are $\vec k$ dependent and unequal in general.
 It will be seen that these small deviations influence the superconducting gap equation symmetries.  
\subsection{Coupled LiC${_6}$ Dispersion Relations }
Li-C hopping  adds a perturbation term to the shrunken graphene Hamiltonian. Obtaining exact dispersions from Eq.~\ref{eq:matrix form- full-sch} is very challenging, so perturbation theory is applied to obtain approximate solutions, as presented in Appendix C.  
However, to get some insight into effects of the coupling, Eq.~\ref{eq:matrix form- full-sch} can be solved exactly at the $\Gamma$ point.
At $\vec k$=0 only the isolated Li band, $E_{Li,0}(0)$ and the lowest valance band, $E_{sh,6}(0)$, are mutually affected. The energies of these bands are, with $E_0(0)\equiv E_{+}$, $E_6(0)\equiv E_{-}$, 
\begin{eqnarray}
{E_{\pm}(0)} = \frac{1}{2}\left( E_{Li,0}(0) + E_{sh,6}(0) \right) \pm 
   \sqrt {{{\frac{1}{4}\left[ E_{Li,0}(0) -E_{sh,6}(0)  \right]}^2} + 6(t_{1}^{LiC})^2}
\end{eqnarray}
and other shrunk graphene bands given by (supplementary)
 Eq.~B.5 remain unchanged. Comparing the fit results from DFT to these equations suggests that $t_{1}^{Li-C}$ is in the 0.3-0.5 eV range, and other next neighbor hopping from Li atoms to C sites are negligible.

There are two critical points in the pure graphene band structure which are affected by decoration and become important: the  charge neutrality  Dirac points folded at the $\Gamma$ point, and the van Hove singularity at the $M$ point.
We define a hopping integral symmetry breaking index, $w_t=\frac{t_1^{'}}{t_{1}}\neq 1$ indicates the
degree of symmetry breaking.  
The difference in Li and C on-site energies can be considered to reflect the amount of doping. 
The Dirac points affected by $w_t$ open a small gap $E_g$ at $\Gamma$, which does not depend 
on $t_{1}^{LiC}$. Depending on doping level,  Li-C hopping  affects the band structure near 
the points that the isolated Li band $E_{Li,0}(\vec{k})$ and uncoupled shrunken 
graphene bands intersect. These impurity effects causes not only changes in energy level but 
alter the density of states.  Superconductivity emerges from pairing of electrons near the 
Fermi energy and it is important to know how 
the density of states at the Fermi energy $N(0)$ changes with decoration.

 \subsection{ Fitting of the seven-band tight binding model to DFT }  
\begin{figure}[ht]
\centering
\includegraphics[width=12cm]{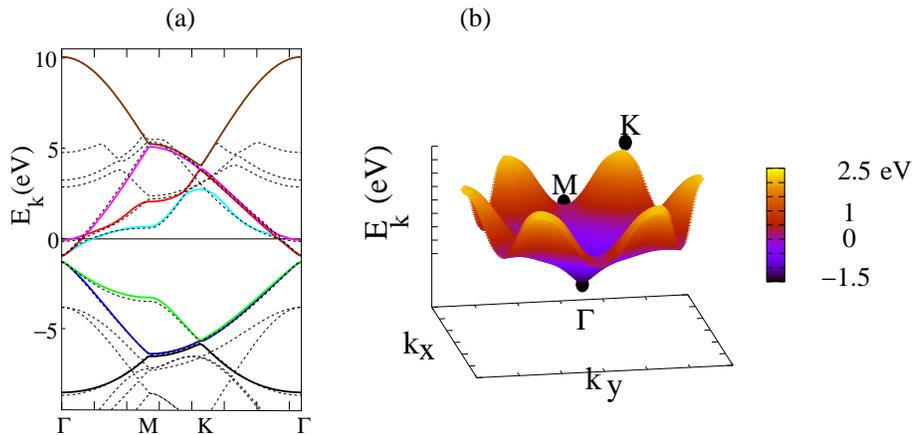}
\caption{(Color online) The left panel provides the band structure of lithium decorated graphene. 
The dashed lines indicate the DFT bands, while the fitted bands are shown in color. The Fermi energy 
set to zero at $\mu_0=0.4 eV$. A small gap, $E_g=0.36 eV$ is opened  at  the $\Gamma$ point around -1.12 eV.
The right panel provides a surface plot of the relatively flat band of LiC$_6$. 
$d$-wave pairing dominates due to electrons in the valleys around saddle points at $M$. }
 \label{figure:fiting-band structure of graphene} 
\end{figure}

The seven band tight binding model of LiC$_6$ was fit to the DFT band structure, with
results illustrated in Fig.~\ref{figure:fiting-band structure of graphene}. In the graphene 
layer shown in Fig.~\ref{figure:graphene decorated by lithium}(a) and (c), 
$A_1$ subsite chosen as central site labeled by $0$  and  $B_1$ subsite in adjacent hexagon considered as second neighbor while just slightly longer than the first neighbors atoms $B_2$ and  $B_3$ in same hexagon,  this neighbor labeled by $n=2$ and so on the next neighbors are labeled. In  Fig.~\ref{figure:graphene decorated by lithium} (a), the big dashed hexagon included up to nine neighbors but for the pristine graphene it is surrounded by five neighbors. 
C-C hopping from $0$-subsite to $n$th neighbor has  been shown by $t^{CC}_{0n}$.
 In-plane Li-Li hopping, $t^{LiLi}_{0m}$ obtained up to $m=4$ neighbors. Li to C hopping 
integrals are very small with respect to those of C-C and Li-Li, so we keep only the near 
neighbor Li-C hopping amplitude.

Since Li is small with respect to alkaline earths such as Ca, the 
pristine band structure is less affected by decoration by lithium than by calcium, 
as can be seen in Fig. 2 of Ref.~[\cite{Profeta}]. 
The fitted hopping amplitudes
 and on-site energies are presented in 
Tables \ref{table:shrunk-hoppping}. Note that by comparing band structure of LiC$_6$ with pristine graphene in ref.\cite{Jung2013}, it is observed that Li decoration only slightly changes 
the pristine graphene band structure. These changes are due to electron transfer from Li 
to graphene, which changes the pristine on site $\epsilon_{pristine}=0$ to   
$\epsilon_A=\epsilon_B=\epsilon_c$.
\section{Superconducting Pairing and States}
\subsection{Bogoliubov-de Gennes Transformation}
LiC$_6$ presents a multiband system in which three bands cross the Fermi level.  We 
presume singlet pairing that can be both intraband and interband in nature.
We adopt a local viewpoint in which pairing occurs between electrons on carbon atoms.  
 Seven hybridized Li-C orbitals, support nine possible bond pairing
amplitudes in real space. Figure~\ref{figure:pairing amplitude} (a) illustrates all the nearest neighbour order parameters possibilities. Leaving the analytical derivation details to supplementary materials Appendices D and E, the quasiparticle energies are obtained by Bogoliubov-de Gennes unitary transformation in the seven band space, 
 \begin{eqnarray}
E_{m,s}^Q (\vec k)= s\left( E_{m}(\vec k)   + \sum_{i = 1}^7 \frac{{\left| \Delta_{mi}(\vec k) \right|}^2}{{E_m}(\vec k) + E_{i}(\vec k) }  \right)  \qquad  s =  \pm 1
\label{eq:quasi-su-spectrum-text}
\end{eqnarray}
in which  $s=1$ is for particles and $s=-1$ for holes, and $E_{m}$ are the normal state eigenvalues.
The  $\vec{k}$-dependent gap $|\Delta_{mi}(\vec k)|^2$ in the spectrum are expressed as
\begin{eqnarray}
\Delta _{mn}(\vec k)
= \sum_{\alpha  = 1}^9 \Omega _{mn}^\alpha (\vec k)\Delta^{\alpha } 
\label{eq:band-orderparameter-matrix-text}
\end{eqnarray}
 in which $m$ and $n$ are band indexes. The band pair order parameter $\Delta _{mn}(\vec k)$ 
denotes pairing between electrons in the $m$-th and $n$-th bands in LiC$_6$. 
Also, $(\Delta^{1},~\Delta^{2},~\Delta^{3})=(\Delta^{''}_{1},\;\Delta^{''}_{2},\;\Delta^{''}_{3});~(\Delta^{4},~\Delta^{5},~\Delta^{6})=(\Delta_{1},\;\Delta_{2},\;\Delta_{3});~
(\Delta^{7},~\Delta^{8},~\Delta^{9})=(\Delta^{'}_{1},\;\Delta^{'}_{2},\;\Delta^{'}_{3})$ are shown in Fig.~\ref{figure:pairing amplitude}(a), and 
\begin{eqnarray}
\Omega^{1}_{ij}(\vec k) &=& \mathscr{C}^{*}_{1 } ({E_i})\mathscr{C}_{4 }({E_j})e^{i\vec k.{{\vec \tau }_1}} + \mathscr{C}^{*}_{4} ({E_i})\mathscr{C}_{1}({E_j})e^{-i\vec k.{\vec \tau }_{1 }}\nonumber\\
\Omega^{2}_{ij}(\vec k) &=& C^{*}_{3 } (E_{i})\mathscr{C}_{6 }({E_j}){e^{i\vec k.{{\vec \tau }_2}}} + \mathscr{C}^{*}_{6} ({E_i})\mathscr{C}_{3 }({E_j})e^{ - i\vec k.{{\vec \tau }_{2}}}\nonumber\\
\Omega^{3}_{ij}(\vec k) &=& \mathscr{C}^{*}_{2 } ({E_i}){\mathscr{C}_{5 }}({E_j}){e^{i\vec k.{{\vec \tau }_3}}} + \mathscr{C}^{*}_{5} ({E_i}){\mathscr{C}_{2 }}({E_j}){e^{ - i\vec k.{{\vec \tau }_3}}}\nonumber \\
\Omega^{4} _{ij}(\vec k) &=& \mathscr{C}^{*}_{2 } ({E_i}){\mathscr{C}_{6}}({E_j}){e^{i\vec k.{{\vec \delta }_1}}} + \mathscr{C}^{*}_{6 } ({E_i}){\mathscr{C}_{2 }}({E_j}){e^{ - i\vec k.{{\vec \delta }_{1}}}} \nonumber\\
\Omega^{5}_{ij}(\vec k) &=& \mathscr{C}^{*}_{1 } ({E_i}){\mathscr{C}_{5 }}({E_j}){e^{i\vec k.{{\vec \delta }_2}}} + \mathscr{C}^{*}_{5 } ({E_i}){\mathscr{C}_{1}}({E_j}){e^{ - i\vec k.{{\vec \delta }_2}}}\nonumber\\
\Omega^{6}_{ij}(\vec k) &=& \mathscr{C}^{*}_{3 } ({E_i}){\mathscr{C}_{4}}({E_j}){e^{i\vec k.{{\vec \delta }_3}}} + \mathscr{C}^{*}_{4} ({E_i}){\mathscr{C}_{3}}({E_j}){e^{ - i\vec k.{{\vec \delta }_3}}}\nonumber\\
\Omega^{7} _{ij}(\vec k) &=& \mathscr{C}^{*}_{3} ({E_i}){\mathscr{C}_{5}}({E_j}){e^{i\vec k.{{\vec \delta }_1}}} + \mathscr{C}^{*}_{5 } ({E_i}){\mathscr{C}_{3}}({E_j}){e^{ - i\vec k.{{\vec \delta }_1}}}\nonumber\\
\Omega^{8} _{ij}(\vec k) &=& \mathscr{C}^{*}_{2} ({E_i}){\mathscr{C}_{4 }}({E_j}){e^{i\vec k.{{\vec \delta }_2}}} + \mathscr{C}^{*}_{4} ({E_i}){\mathscr{C}_{2 }}({E_j}){e^{ - i\vec k.{{\vec \delta }_2}}}\nonumber\\
\Omega^{9} _{ij}(\vec k) &=& \mathscr{C}^{*}_{1 } ({E_i}){\mathscr{C}_{6 }}({E_j}){e^{i\vec k.{{\vec \delta }_3}}} + \mathscr{C}^{*}_{6 } ({E_i}){\mathscr{C}_{1}}({E_j}){e^{ - i\vec k.{{\vec \delta }_3}}}.
\label{eq:band-orderparameter-coef-text}
\end{eqnarray}
where  $\mathscr{C}_{i } ({E_j})$  are Bloch wave coefficients of the $j$-th band. 
Possible order parameter symmetries in Eq.~\ref{eq:band-orderparameter-matrix-text} 
are related to symmetries of Bloch wave functions, through 
$\Omega_{ij}(\vec k)$ functions in 
Eq.~\ref{eq:band-orderparameter-coef-text}. In the limiting case of (folded) six band pristine 
graphene, the symmetry character of different conduction bands along high symmetry 
lines were provided  in  Fig.~\ref{figure:eta-k}. Bloch symmetry character of non-interacting bands specifies the symmetry of the band  order parameter.
\subsection{Superconducting States}
The linearized gap equation, obtained by minimizing the quasiparticle free energy with 
respect to nearest neighbor order parameters, is
\begin{equation}
J_{\beta }\Delta^{\beta } =- \frac{1}{2N}\sum_{\alpha =1}^9 \left[ \sum_{\vec k} 
   \sum_{n = 1}^7 \sum_{i = 1}^7 \frac{\tanh (\frac{{ E_n^Q}}{ 2 k_{B}T})}{E_{n}(\vec k) 
  + E_{i}(\vec k) } \left( \Omega _{ni}^\alpha (\vec k)\Omega _{ni}^{ * \beta }(\vec k) 
  + \Omega _{ni}^\beta (\vec k)\Omega _{ni}^{ * \alpha }(\vec k) \right) \right] \Delta^{\alpha }
 \equiv - \sum_{\alpha  = 1}^9 \Gamma _{\beta \alpha } \Delta^{\alpha }.
\label{eq:gap-Equation-text}
\end{equation}
This equation can be written in matrix form as   
\begin{eqnarray}
\left[ \begin{array}{*{20}{c}}
A_{3 \times 3}&B_{3 \times 3}&B_{3 \times 3}\\
B_{3 \times 3}&C_{3 \times 3}&D_{3 \times 3}\\
B_{3 \times 3}&D_{3 \times 3}&C_{3 \times 3}
\end{array} \right]\left( \begin{array}{*{20}{c}}
g_{1}V_{1}\\
g_{0}V_{2}\\
g_{0}V_{3}
\end{array} \right) =- \left( \begin{array}{*{20}{c}}
V_{1}\\
V_{2}\\
V_{3}
\end{array} \right)
\label{eq:matrix-form-gap-eq-text}
\end{eqnarray}
where
\begin{eqnarray}
A_{3 \times 3} = \left[ \begin{array}{*{20}{c}}
\Gamma _{11}&\Gamma _{12}&\Gamma _{12}\\
\Gamma _{12}&\Gamma _{11}&\Gamma _{12}\\
\Gamma _{12}&\Gamma _{12}&\Gamma _{11}
\end{array} \right],\;    C_{3 \times 3} = \left[ \begin{array}{*{20}{c}}
\Gamma _{44}&\Gamma _{45}&\Gamma _{45}\\
\Gamma _{45}&\Gamma _{44}&\Gamma _{45}\\
\Gamma _{45}&\Gamma _{45}&\Gamma _{44}
\end{array} \right],\;\;
B_{3 \times 3} = \left[ \begin{array}{*{20}{c}}
\Gamma_{14}&\Gamma_{15}&\Gamma_{15}\\
\Gamma_{15}&\Gamma_{14}&\Gamma_{15}\\
\Gamma_{15}&\Gamma_{15}&\Gamma_{14}
\end{array} \right],\;\;    D_{3 \times 3} = \left[ \begin{array}{*{20}{c}}
\Gamma_{47}&\Gamma_{48}&\Gamma_{48}\\
\Gamma_{48}&\Gamma_{47}&\Gamma_{48}\\
\Gamma_{48}&\Gamma_{48}&\Gamma_{47}
\end{array} \right]
\label{eq:matrix-gap-elements-text}
 \end{eqnarray}
and $g_1V_1=\left(\Delta^{''} _{1}~\Delta^{''} _{2 }~\Delta^{''} _{3} \right)^T$,
 $g_0V_2=\left(\Delta _{1 }~\Delta _{2 }~\Delta _{3 } \right)^T$ and 
 $g_0V_3=\left(\Delta^{'} _{1 }~\Delta^{'} _{2 }~\Delta^{'} _{3 } \right)^T$;  
the subscripts $<ij>$  has been dropped for brevity.
 The $A_{3 \times 3}$, $B_{3 \times 3}$, $C_{3 \times 3}$, and $D_{3 \times 3}$  matrices, given by 
 Eq.~\ref{eq:matrix-gap-elements-text},
  have identical structures, 
hence they share eigenvectors: 
$V_{s}^{T}=(1\;\;1\;\;1)$, $V_{d_{xy}}^{T}=(1\;\;-1\;\;0)$, and 
$V_{d_{x^2-y^2}}^{T}=(1\;\;1\;\;-2)$, where the latter two are degenerate. Their eigenvalues,
in obvious notation, are 
\begin{eqnarray}
{{a}_s}& =&{\Gamma _{11}} + 2{\Gamma _{12}}\;,\;\;  
{{b}_s}={\Gamma _{14}} + 2{\Gamma _{15}}\;,\;\; 
{{c}_s}={\Gamma _{44}} + 2{\Gamma _{45}}\;,\;\; 
{{d}_s}= {\Gamma _{47}} + 2{\Gamma _{48}}\nonumber\\
{a_d} &=&{\Gamma _{11}} - \;\;{\Gamma _{12}}\;,\;\; {b_d} ={\Gamma _{14}} - \;\;{\Gamma _{15}}
\;,\;\;  {c_d} ={\Gamma _{44}} - \;{\Gamma _{45}}\;,\;\;   {d_d} ={\Gamma _{47}} -\; {\Gamma _{48}}.
\label{eq:gapequation-matrix-unhermit-elements-text}
\end{eqnarray}
For folded six band pure graphene $g_0=g_1$, the  Bloch wave coefficients appearing
in Eq.~\ref{eq:band-orderparameter-coef-text} can be replaced from 
Eq.~B.7
 to show that $\Omega_{ij}^{1}(\vec k)=\Omega_{ji}^{4}(\vec k)=\Omega_{ij}^{7}(\vec k)$ and similarly relations for other elements, hence $C_{3 \times 3}=A_{3 \times 3}$ and $D_{3 \times 3}=B_{3 \times 3}$.  Eq.~\ref{eq:matrix-form-gap-eq-text}
 takes the more symmetric form
\begin{eqnarray}
\left[ \begin{array}{*{20}{c}}
A_{3 \times 3}&B_{3 \times 3}&B_{3 \times 3}\\
B_{3 \times 3}&A_{3 \times 3}&B_{3 \times 3}\\
B_{3 \times 3}&B_{3 \times 3}&A_{3 \times 3}
\end{array} \right]\left( \begin{array}{*{20}{c}}
V_{1}\\
V_{2}\\
V_{3}
\end{array} \right) =-\frac{1}{g_0}\left( \begin{array}{*{20}{c}}
V_{1}\\
V_{2}\\
V_{3}
\end{array} \right)
\label{eq:matrix-form-gap-eq-pristine-text}
\end{eqnarray}
For the case $V_1=V_2=V_3=V_{sy}$ where $sy$ subscripts indicates each of the $s$, 
$d_{xy}$  or $d_{x^2-y^2}$ symmetry, the six band gap 
Eq.~\ref{eq:matrix-form-gap-eq-pristine-text} reduces to 
$\left(A+2B\right)V_{sy}=-\frac{1}{g_0}V_{sy}$, i.e. the linearized gap equation of the
two band model of pristine graphene in Ref. [\cite{Blackschaffer2007}]. 
These three solutions preserve symmetry of the two band unit cell as illustrated in 
Fig.~\ref{figure:pairing amplitude}~(b), (c). In addition to these three  states, 
there are six more non-orthogonal solutions  $\Phi_{0n}=(V_{sy}~0~-V_{sy})$ and 
$\Phi_{1n}=(V_{sy}~-V_{sy}~~0)$ that break symmetries of pristine graphene two band model. 
Inserting these 
solutions into Eq.~\ref{eq:matrix-form-gap-eq-pristine-text} leads to a new
 two band gap equation,  $\left(A-B\right)V_{sy}=-\frac{1}{g_0}V_{sy}$, which is
unphysical because of an unreachably high energy pairing potential $g_0$. 
In the following section the superconducting gap equation has been solved for $LiC_6$ 
and it is demonstrated how Li-C coupling influences superconducting phases.

\subsection{Nine Superconducting Phases}
Self-consistent solutions of the gap equation Eq.~\ref{eq:matrix-form-gap-eq-text} 
can be obtained analytically.
There are three superconducting states with island character (discussed in more 
detail below)  that can be expressed
in compact form as
\begin{eqnarray}
[\Phi_{n}]^{T}=[0\quad V_{{sy}}\quad -V_{sy}], ~~~J_{sy}^{0}= c_{sy} - d_{sy}
\label{eq:gap-general-function-island-text}
\end{eqnarray}
 where  $V_{sy}$ refers to one of the $V_s$, $V_{d_{xy}}$ or $V_{d_{x^2-y^2}}$-wave symmetries.  Pairing in these phases cannot propagate, as may be pictured in Fig.~\ref{figure:dxy9}.
The other   six superconducting states of 
Eq.~\ref{eq:matrix-form-gap-eq-text} have the explicit form
\begin{eqnarray}
[\Phi_{n}]^{T}=[\alpha_{sy}^{\pm}V_{{sy}}\quad V_{{sy}}\quad V_{sy}]
\label{eq:gap-general-function-text}
\end{eqnarray}
  corresponding to the interaction potential is $g_0=\frac{1}{J_{sy}}$ wherein
 \begin{eqnarray} 
\alpha _{sy}^{\pm} =  \frac{J_{sy}^{\pm} - c_{sy} - d_{sy}}{b_{sy}},~~
J_{sy}^{\pm} =\frac{1}{2}\left({\kappa {a_{sy}} + {c_{sy}} + {d_{sy}} 
             \pm \sqrt {8\kappa b_{sy}^2 + {{\left[ {{c_{sy}} + {d_{sy}} 
             - \kappa {a_{sy}}} \right]}^2}} } \right)
\label{eq:quad-eigenvalue-text}
\end{eqnarray}
In these expressions  $J_{d(s)}^{\pm}$ and  $\alpha_{d(s)}^{\pm}$ are obtained from 
Eq.~\ref{eq:quad-eigenvalue-text} by substituting $a_{sy},$  $b_{sy},$  
$c_{sy},$ and $d_{sy}$ by $a_{d(s)},$  $b_{d(s)},$  $c_{d(s)},$ and 
$d_{d(s)}$ respectively.

By comparing  the gap equations introduced in Eq.~\ref{eq:matrix-form-gap-eq-text}  and  Eq.~\ref{eq:matrix-form-gap-eq-pristine-text} the gap equation symmetry reduction of decorated graphene with 
respect to folded but pristine graphene becomes clear.  
This symmetry reduction results in an $\alpha_{sy}$ coefficient  appearing in the pairing 
amplitudes of stretched bonds as shown in Eq.~\ref{eq:gap-general-function-text} and 
Fig.~\ref{figure:pairing amplitude}. We refer to these symmetry reduction phases as ``distorted phases''.

The six bands of pristine graphene  support nine pairing amplitudes while in the two band model  
there are three possible pairing amplitudes along three different  bonds. These two notions 
can be mapped onto each other only if $\alpha_{sy}=1$ as illustrated  in 
Fig.~\ref{figure:pairing amplitude}(b),(c).  Therefor the three island superconducting phases   
given by Eq.~\ref{eq:gap-general-function-island-text} in the special case of pristine cannot 
be mapped onto the two band model. These three phases are unphysical even in the case of 
decorated graphene because the Cooper pairs in these phases require a large pairing potential. 
In the special case of pristine graphene in which $\kappa=1$, $a_{sy}=c_{cy}$ and $b_{sy}=d_{cy}$ 
from Eq.~\ref{eq:quad-eigenvalue-text}, and it follows that if $b_{sy}>0$ then $\alpha_{sy}^{+}=1$ 
and $\alpha_{sy}^{-}=-2$. Also $g_{0}^{+}<g_{0}^{-}$  so in this case $(+)$ sign preserves 
the two band model while the $(-)$ sign phases are unphysical. 
Numerical calculation shows that $b_{sy}^{+}>0$.
 These superconducting states can be categorized into three groups according to their 
corresponding pairing potential.
 \subparagraph{3 electron pairing states with  island character and very high pairing potential; (unphysical solutions)}
\begin{eqnarray}
\Phi_{p_x}&=\frac{1}{2}&[(0~0~0)~(1-1~0)(-1~1~0)]^T,~~~g_0=\frac{1}{ c_d-d_d}, \nonumber \\
\Phi_{p_y}&=\frac{1}{2\sqrt{3}}&[(0~0~0)~(1~1-2)(-1-1~2)]^T,~g_0=\frac{1}{ c_d-d_d}, \nonumber \\
\Phi_{f}&=\frac{1}{\sqrt{6}}&[(0~0~0)~(1~1~1)(-1-1-1)]^T,~g_0=\frac{1}{ c_s-d_s}.
\label{eq:gap-eigenstates-1}
\end{eqnarray}
\subparagraph{3 states with higher electron pairing potential; (unphysical solutions)}
\begin{eqnarray}
\Phi_{d_{{x^2} - {y^2}}}^{{-}} &=& \left (6 (\alpha^{-}_{d})^2+12\right )^{-\frac{1}{2}}{\left[ {\begin{array}{*{20}{c}}
{  {\alpha _{d}^{-}}\left( {1\quad 1\;-2} \right)}&{\left( { 1\; 1\;-2} \right)}&{\left( { 1\; 1\;-2} \right)}
\end{array}} \right]^T},~~g_{0}=\frac{1}{J_{d}^{-}}\nonumber \\
\Phi_{d_{xy}}^{-} &=& \left (2 (\alpha^{-}_{d})^2+\;\;4\right )^{-\frac{1}{2}}{\left[ {\begin{array}{*{20}{c}}
{  {\alpha _{d}^{-}}\left( {1-1\;\quad 0} \right)}&{\left( {1\;-1\;0} \right)}&{\left( {1\;-1\; 0} \right)}
\end{array}} \right]^T},~~g_{0}=\frac{1}{J_{d}^{-}}\nonumber \\
\Phi_{s}^{-} &=& \left (3(\alpha^{-}_{s})^2+~~6\right )^{-\frac{1}{2}}{\left[ {\begin{array}{*{20}{c}}
{  {\alpha _{s}^{-}}\left( {1\quad 1\quad 1} \right)}&{\left( {1\quad 1\quad 1} \right)}&{\left( {1\quad 1\quad 1} \right)}
\end{array}} \right]^T,~~g_{0}=\frac{1}{J_{s}^{-}}}
\label{eq:phi-minus}
\end{eqnarray}
\subparagraph{3 states with lower electron pairing potential; (physical solutions)}
\begin{eqnarray}
\Phi_{d_{{x^2} - {y^2}}}^{{+}} &=& \left (6 (\alpha^{+}_{d})^2+12\right )^{-\frac{1}{2}}{\left[ {\begin{array}{*{20}{c}}
{  {\alpha _{d}^{+}}\left( {1\quad 1\;-2} \right)}&{\left( { 1\; 1\;-2} \right)}&{\left( { 1\; 1\;-2} \right)}
\end{array}} \right]^T},~~g_{0}=\frac{1}{J_{d}^{+}}\nonumber \\
\Phi_{d_{xy}}^{+} &=& \left (2 (\alpha^{+}_{d})^2+\;\;4\right )^{-\frac{1}{2}}{\left[ {\begin{array}{*{20}{c}}
{  {\alpha _{d}^{+}}\left( {1-1\;\quad 0} \right)}&{\left( {1\;-1\;0} \right)}&{\left( {1\;-1\; 0} \right)}
\end{array}} \right]^T},~~g_{0}=\frac{1}{J_{d}^{+}}\nonumber \\
\Phi_{s}^{+} &=& \left (3(\alpha^{+}_{s})^2+~~6\right )^{-\frac{1}{2}}{\left[ {\begin{array}{*{20}{c}}
{  {\alpha _{s}^{+}}\left( {1\quad 1\quad 1} \right)}&{\left( {1\quad 1\quad 1} \right)}&{\left( {1\quad 1\quad 1} \right)}
\end{array}} \right]^T},~~g_{0}=\frac{1}{J_{s}^{+}}.
\label{eq:phi-plus}
\end{eqnarray}
All states are orthogonal except those with same subscript, viz.  ${\Phi_{s}^{{-}}}$ 
and ${\Phi_{s}^{{+}}}$. Such solutions are orthogonal if $\kappa=1$, {\it i.e.} $g_{1}=g_{0}$. 
Only for this case the matrix gap equation becomes Hermitian, then band order 
parameters takes the following form in terms of the band Green function and $g_{0}$, 
\begin{eqnarray}
\Delta_{ij}(\vec k)=g_{0}\langle d^{\uparrow}_{i}(\vec k)d^{\downarrow}_{j}(\vec k)\rangle.
\label{eq:order-green}
\end{eqnarray}
Here $\hat{d}^{\sigma}_{i}(\vec k)=\sum^{7}_{m=1}{\mathscr{C}}^{*}_{m} (E_{i}(\vec k)) 
 \hat{c}^{\sigma}_{m}(\vec k)$ annihilates an electron with spin $\sigma$ in the $i$th band with 
energy $E_{i}(\vec k)$. Although it is assumed that $g_1=g_0$ but deviation from pristine 
leads to distortion of Green's functions 
$<\hat{c_{i\alpha}^{\sigma}}\hat{c_{j\beta}^{\sigma}}>$ along different bonds. 

Phases $\Phi_{d_{{x^2} - {y^2}}}^{{-}}$ and $\Phi_{d_{xy}}^{-}$ are degenerate with 
eigenvalue $J_{d_{x^2-y^2}}^{-}=J_{d_{xy}}^{-}=J_{d}^{-}$, and similarly 
$\Phi_{d_{{x^2} - {y^2}}}^{{+}}$ and $\Phi_{d_{xy}}^{+}$ with eigenvalue $J_{d}^{+}$. 
For Li decorated graphene, numerical calculation shows 
$J_{sy}^{+}>J_{sy}^{-}$, so $g_{0}$ in the $(+)$ states is lower 
than $g_{0}$ in the $(-)$ states hence pairing in this modes are dominant.  
From Eqs.~\ref{eq:phi-minus} and \ref{eq:phi-plus} we observe that probability amplitudes for pairing 
on different bonds in real space differ for the various states. For the long C-C bonds 
the probability is proportional to $(\alpha^{\pm}_{sy})^2$ while for the others is unity. Numerical results are shown inFig.~\ref{figure:g0-Tc}. 
\section{ Discussion And Relation To Previous Work }
The possibility of a superconductivity state in metal decorated graphene has been suggested theoretically by a few groups.\cite{Uchoa,Blackschaffer2007,Profeta} Some have suggested phonon-mediated superconductivity in single layer graphene.  Most prominently, Profeta {\it et al.}\cite{Profeta} calculated on the basis of density functional theory for superconductors that decoration by electron donating atoms such as Ca and Li will make 
single layer graphene superconducting, up to 8K for the case of Li. The {\it ab initio} anisotropic Migdal-Eliashberg formalism was used by Zheng and Margine\cite{Zheng2016}, who predicted  a single anisotropic superconducting gap  with critical temperature $T_c=5.1-7.6 K$, in surprisingly good agreement with experimental reported superconductivity around 6K in $LiC_6$.\cite{{Ludbrook2015}}
 
Using a phenomenological  microscopic Hamiltonian in a nearest-neighbor tight-binding approximation, possible superconducting  phases of pristine  graphene have been discussed by Uchoa and Castro-Neto\cite{Uchoa} and also by Black-Schaffer and Doniach\cite{Blackschaffer2007}.  The possibility of  a singlet $p+ip$ phase pairing near the Dirac points between nearest neighbors subsites were suggested by Uchoa and Castro-Neto\cite{Uchoa}. They worked in terms of a plasmon mediated mechanism for metal coated graphene, and discussed the conditions under which attractive electron-electron interaction can be mediated by plasmons.
  
Singlet superconducting gap phases of pristine graphene have been proposed and discussed 
by Black-Schaffer and Doniach.\cite{Blackschaffer2007} 
 For the nearest neighbors pairing amplitudes $\Delta_{<iAjB>}=\Delta_{iA,iA+\vec{\delta_j}}$ where $\vec{\delta_j}$ are the vectors that connects the $iA$ site to its three nearest neighbors, it was observed that
there are three states that minimize the free energy in various regimes of the parameters, which here have been denoted by $V_s=(1,1,1)^T$, $V_{d_{x^2-y^2}}=(2,-1,-1)^T$, and $V_{d_{xy}}=(0,-1,1)^T$.  Pairing symmetries  
$d_{xy}$ and $d_{x^2-y^2}$ are degenerate, and only the linear  combination of   
${d_{x^2-y^2}}+i{d_{xy}}\equiv d+id$ preserves the graphene band symmetry. 
Depending on the position of the Fermi energy with respect to Dirac points, $d+id$ or $s$ 
states tend to dominate. Their numerical calculation showed that $d$-wave solutions will always be favored for electron or hole doping in the regime $0<\bar{n}_c<0.4$ where doping is defined by $\bar{n}_\alpha=<\hat{c}_{i\alpha}^{\dag}\hat{c}_{i\alpha}>-1$. In this regime, superconductivity can emerge from electronic correlation effects. Near the van Hove singularity at the  saddle point $M$  corresponding  to $3/8$ and $5/8$ fillings i.e. $\bar{n}_c=0.25$, it was suggested
that chiral $d+id$ superconductivity, which breaks time-reversal symmetry, can be stabilized. In this regime $d$-wave superconductivity may arise from repulsive electron-electron interaction\cite{Nandkishore2014}.

Although doping by a gate voltage 
is normally considered to change only the chemical potential but not the band structure,  
gating cannot be expected to push the Fermi energy to the van Hove singularity without
altering the band dispersion. The most
likely way to do this is by decoration with electropositive atoms, which has been our focus.
We note that doping is essential, when graphene decorated, in addition to the expected charge migration from the decorating atoms to the graphene sheet, it is then  necessary the interlayer state
is partially  occupied   to induce superconductivity as happens in GICs. Hybridization of interlayer $s$-band and graphene $\pi$ bands changes the graphene band structure. 
The $s$ orbitals of Ca
have more overlap with C orbitals than Li and lead to stronger and longer range interactions as
well as increasing the doping level, effects that become detrimental to superconductivity. For this 
reason our emphasis here  is on the Li decorated graphene.

We review some of our main points.
When graphene is decorated by Li, electron transfer from Li atoms to C 
contracts the Li-C distance and
reduces the C-C bond lengths in the Li-centered hexagon. In this kekul\'e -type structure, 
hopping amplitude symmetries of all C-C neighbors are broken (our `` shrunken graphene'').
This model allows study of multiband effects on the superconducting phase diagram.
To gain insight into our model,  solutions of superconducting gap equation in both cases of folded bands otherwise pristine C$_6$  and the usual two band model of C$_2$ were compared. 
These two viewpoints coincide if the same pairing paradigms are considered.
For pristine graphene with its two site cell, in real space picture electrons can pair  
with near neighbors in three inequivalent directions, 
$\Delta_{i,i+\vec{\delta}}=V_{sy}=(\Delta_1~\Delta_2~\Delta_3)^T$ which must respect honeycomb 
symmetries.  The $V_{sy}$ quantities are the three vectors that belong to the irreducible 
representation of crystal point group D$_{6h}$ i.e. $V_{sy}^T=$ (1,1,1), (-1,1,0) and (2,-1,-1) 
for which the $sy$ subscript stands for symmetries $s$ , $d_{xy}$ and $d_{x^2-y^2}$. 
Permutation of $s$-wave solution (1,1,1) along three different bonds  constructs just one state 
while permutation of d$_{xy}$ solution (-1,1,0) up to a minus sign constructs two 
nonorthogonal linear independent states viz. (-1,1,0) and  (-1,0,1) which  orthogonal  linear combination of 
them are d$_{xy}^T=$(-1,~1,~0) and $d_{x^2-y^2}^T=$ (2~-1~-1). 

\begin{figure}[ht]
\centering
\includegraphics[width=13cm]{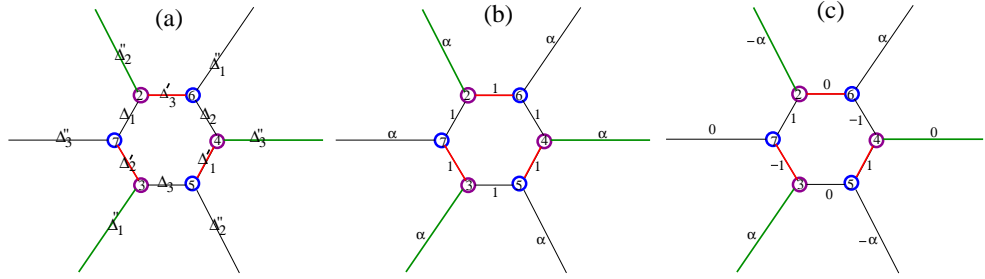}
\caption{(Color online) 
(a) Designation of the pairing amplitudes considered in this study, 
which cover all nearest neighbor pairing possibilities denoted by $\Delta_{n{<ij>}}$, 
$\Delta^{'}_{n{<ij>}}$ and $\Delta^{''}_{n{<ij>}}$ where subscript $<ij>$  has been dropped 
for brevity. (b) shows the pairing amplitude for $\Phi^{+}_{S}$  phase with $\alpha\approx 0.6$ and 
for $\Phi^{-}_{S}$ phase  with $\alpha\approx -3.4$. 
Both phases broken two band graphene symmetry as can 
be seen by comparing symmetries along different bonds in seven atoms unit cell and two bands unit cell where its Bravais lattice points are labeled by 5, 6 and 7. 
 (c) shows the pairing amplitude $\Phi^{+}_{d_{xy}}$ where 
$\alpha\approx 1$ and $\Phi^{-}_{d_{xy}}$ where $\alpha\approx -2$. The first phase 
approximately preserves two band graphene symmetry while the others arise from broken symmetry.   
  }
 \label{figure:pairing amplitude} 
\end{figure}

\begin{figure}[ht]
\centering
\includegraphics[width=8cm]{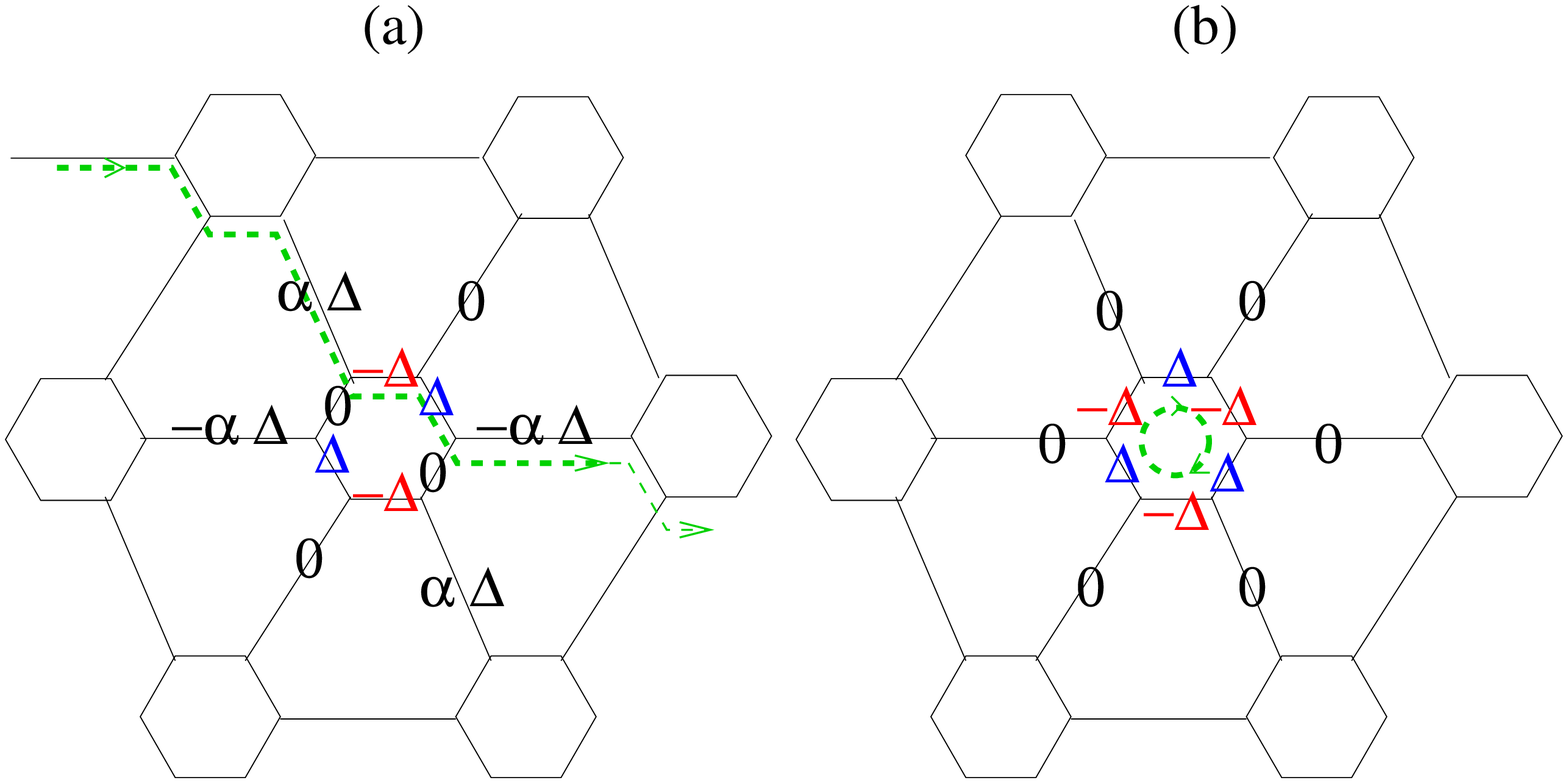}
\caption{(Color online) Schematic diagram illustrating Cooper pair propagation for 
(a) $\Phi^{J}_{xy}$ along a chain as shown by the green dashed line and arrow,  and 
(b) localized ``island'' pairs for $\Phi_{f}$. }
 \label{figure:dxy9} 
\end{figure}

A similar procedure again can 
be applied to pristine graphene but now in enlarged six site unit cell. Unit cell of C$_6$ 
includes six carbon subsites  and  nine different bonds that support nine possible nearest 
neighbor bond pairing amplitudes as illustrated in Fig.~\ref{figure:pairing amplitude} 
and denoted them  by 
$\Phi_{sy}^{T}=[(\Delta^{''}_{1},~\Delta^{''}_{2}~\Delta^{''}_{3})(\Delta^{1},~\Delta^{2},~\Delta^{3})(\Delta^{'}_{1},~\Delta^{'}_{2}~\Delta^{'}_{3})]$. 
The gap equation is a $9\times9$ matrix equation given by 
Eq.~\ref{eq:matrix-form-gap-eq-pristine-text}.  The folded  bands supercell 
include three vertices numbered  5, 6, 7,  and nine bonds as shown in 
Fig.~\ref{figure:pairing amplitude}(a). There are nine orthogonal solutions that preserve 
symmetries of this supercell. One of these configurations has s-wave symmetry 
$(1,1,1,1,1,1,1,1,1)$ the other eight solutions  are constructed by all possible 
permutations of (-1,1,0) along these bonds that preserve our supercell symmetry.
There are only three solutions which can preserve symmetry of both two and six atoms cells 
simultaneously which they are of the form 
$\Phi_{sy}^{+}=(V_{sy}~V_{sy}~V_{sy})^T$  as illustrated in Fig.~\ref{figure:pairing amplitude}.
For these  solutions, the folded $9\times 9$ gap equation reduces to $3\times 3$  gap equations 
of ordinary pristine graphene. 
The Cooper pair formation energy for these three modes are significantly less than the other 
six phases which 
are not reducible to the two band model. 

In fact reduction of symmetry leads to increasing of the system free energy.
After the orthogonalization procedure, one obtains three solutions 
$\Phi_{f}$, $\Phi_{p_x}$ and $\Phi_{p_y}$, of the form
$\Phi_{sy}^{0}=(0~~~V_{sy}~-V_{sy})^T$.
These phases have been designated as island phases, as illustrated in 
Fig.~\ref{figure:dxy9}(b) for 
$\Phi_{f}$, within which a pairing amplitude is localized within island hexagons 
and cannot propagate. 
For these island phases, numerical calculation of the electron pair potential energy $g_0$ 
shows that $g_0$ is large.
This kind of solutions is a consequence of the six atom basis and does not appear for the 
two atom basis. Also, there are three solutions of the form 
$\Phi_{sy}^{-}=(-2V_{sy}~V_{sy}~V_{sy})$  which also break symmetry of two atom cell. 
For these reasons, in association with the normal state band structure of graphene, we concentrate on superconductivity in the three $\Phi_{sy}^{+}$ symmetry phases.

For pristine graphene C$_2$, two normal  bands are $E_{\pm}=\pm t_1 |\eta_0|$ which  fold to six branches in mini-BZ of C$_6$ i.e.  $E_{\gamma}^{\pm}=\pm t_1 |\eta_0|$, $E_{\beta}^{\pm}=\pm t_1 |\eta_1|$ 
and $E_{\alpha}^{\pm}=\pm t_1 |\eta_2|$    as shown in
{figure:eta-k}, also  
Bloch-wave symmetry character of each branch has been distinguished. 
The Bloch coefficients of the branch  labeled by $\gamma$  are of $s$-wave  character, 
$C_{A_i}=(1,1,1)$ and for those labeled as $\alpha$  and $\beta$  are of the form 
$d\pm id$ type, {\it i.e.} $C_{A_i}=(1,e^{\pm i\frac{i2\pi}{3}},e^{\pm i\frac{i4\pi}{3}})$. 
Based on Bloch wave character of these  branches one can obtain the dominant superconducting  
phases of pristine graphene in various doping regimes. $d$-wave pairing emerges from the $d$-wave 
branches of the folded band structure $E_{\alpha}^{\pm}$ and $E_{\beta}^{\pm}$, while 
$s$-wave pairing arises from the $s$-wave branch $E_{\gamma}^{\pm}$. 
For folded but otherwise pristine graphene, Fig.~\ref{figure:eta-k} illustrates that the 
lowest conduction band, weakly dispersive  along 
$\Gamma\rightarrow M$, is responsible for dominant
singlet superconductivity in chiral $d\pm id$ symmetry. Upon  
electron doping to the critical vHs at $\bar{n}_c=0.25$, the pairing potential $g_0$ in the 
$d\pm id$ phase decreases, beyond which  density of states decreases. $g_0$ increases 
until a second critical value of
doping $\bar{n}_c=0.4$ at which a phase transition to $s$-wave pairing occurs. 
Bloch states in higher conduction bands include  combinations of $s$ and $f$ 
symmetries that favor extended $s$ wave pairing. The multiband character is responsible 
for stabilizing singlet $s$ superconductivity at high electron or hole doping.

To understand how superconducting phases of graphene can be affected by decoration by Li, 
one can compare the LiC$_6$  gap solutions with those of folded bands C$_6$ at the same doping. 
Numerical results for pristine graphene gap equation performed in the nearest neighbor 
approximation in Ref.~\cite{Blackschaffer2007} have been extended by applying  
a more accurate tight binding model fit to the DFT band structure of pristine 
graphene\cite{Jung2013}. Although  a quantum critical point for zero doping reported by 
Black-Schaffer and Doniach\cite{Blackschaffer2007} at dimensionless coupling 
$\frac{g_0}{t}=1.91$ which $d$- and $s$-wave solutions are degenerate. In the more realistic 
tight binding model we applied, this degeneracy is not observed at the $\Gamma$ point, and 
the $d$-wave solution is dominant. This difference may be consequence of particle-hole 
symmetry breaking of valence and conduction bands. Also the van Hove singularity at the M point 
is moved from 0.25 doping for nearest neighbor hopping to 0.16 doping in the accurate model.
The phase transition from $d$-wave to $s$-wave is shifted to 0.35 doping instead of the 0.4 
doping reported for nearest neighbor hopping.\cite{Blackschaffer2007}.  
Numerical calculations for this more detailed model are illustrated in
 Fig.~\ref{figure:g0-n}.

When graphene is decorated by Li, around 0.68 electron per lithium atom transfers to neighboring
C sites, viz. $\bar{n}_c = 0.11$, and 
the Dirac points folded to $\Gamma$ move to -1.52 eV. Symmetry breaking of the hopping  
partially removes degeneracies of band structure of pristine graphene, which leads to creation 
of the small gap at $\Gamma$, with energy  $E_g=2|t_1-t_{1}^{'}|=0.36eV$. Also two of four-fold 
degeneracies  between valence and conduction bands at the Dirac points are removed. 
Compression between band structure of decorated graphene and folded pristine graphene at 
the same doping shows that hybridization of the Li $s$ band and C $\pi$ band is small. This means 
nearest neighbor Li-C hopping is in the range $t_{1}^{LiC}\sim0.3-0.5$, and further hoppings 
are negligible.  

Li decoration of graphene changes not only the band structure but also the Bloch wave 
coefficients from those of pristine graphene. While pristine graphene Bloch wave coefficients 
have pure $s$- or $d$-wave character and their magnitudes are $\vec{k}$-independent. In 
the case of LiC$_6$ they become mixed and vary with $\vec{k}$, hence gap equation symmetry 
is reduced. Because of this symmetry reduction,
for the longer C-C bonds, a new coefficient $\alpha_{sy}$  
appears in the pairing amplitudes. In terms of this coefficient we have classified  
superconducting phase symmetries into three groups. 
Eqs.~\ref{eq:gap-eigenstates-1}, \ref{eq:phi-minus}, and \ref{eq:phi-plus} present all nine 
possible pairing phases of LiC$_6$. There are three categories of solutions which have not
appeared in complete form in the literature. The total of nine phases arise from spatial, and 
therefore hopping parameter, symmetry breaking.

In the first category 
$\Phi_{f}$, $\Phi_{p_x}$ and $\Phi_{p_y}$, there is $\alpha_{sy}=0$ identical to that of 
folded pristine C$_6$.  
For the second category, $\alpha_{sy}$ (denoted by $\alpha^{-}$) is negative, 
in the case of pristine $\alpha^{-}=-2$  as discussed. 
These three phases break the two site cell symmetry, and numerical calculation shows 
that the pairing potential $g_0$ must be large to realize these phases.   
For the last category $\alpha^{+}$ is positive. Three phases which correspond to $\alpha^{+}>0$ 
include $\Phi^{+}_{d_{x^2 - y^2}}$, $\Phi^{+}_{d_{xy}}$, and $\Phi^{+}_{s}$, and these have the 
lowest pairing potentials with respect to the other six phases. 

\begin{figure}[ht]
\centering
\includegraphics[width=12cm]{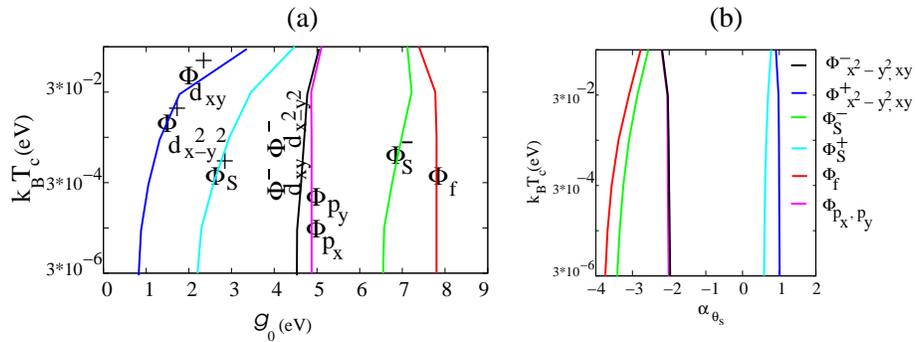}
\caption{(Color on line) (a) This phase diagram illustrates the relation between $T_{c}$ 
and the pairing potential $g_{0}$ for Li$C_6$ in which $\mu_0=0$.  
Panel (b) shows $T_{c}$ in terms of $\alpha_{sy}$  }
 \label{figure:g0-Tc} 
\end{figure}

\begin{figure}[ht]
\centering
\includegraphics[width=8cm]{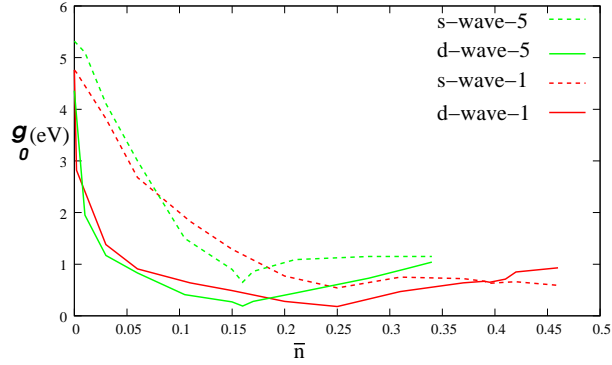}
\caption{(Color online) shows cooper pair interaction $g_{0}$ in terms of doping $\bar{n}$ for d and s-wave phases for pristine graphene at  T= 0.1K. The solid (dashed) red line  indicates d- wave (s- wave) pairing interaction in first nearest neighbor hopping $t_1=2.5 eV$ and similarly green line for accurate tight binding model can fit on DFT. For red line at the charge neutrality  s- and d- wave are degenerate with $g_0=4.76$ while for full approximation they are not degenerate.  }
\label{figure:g0-n} 
\end{figure}

In the limiting case of folded six band pristine graphene $\alpha_{d_{x^2-y^2}}^{+}$, 
$\alpha_{d_{xy}}^{+}$,  and $\alpha^{+}_{s}$  are all equal to unity, which maps the results to
the two-band symmetries as it should.  
But when Li decorated, depending on doping  strength viz. $w_t$ and $t_{1}^{LiC}$ these 
coefficients $\alpha^{+}_{sy}$ no longer remain 
unity.  The pairing amplitude distortion along longer C-C bonds $\alpha^{+}$, for s-wave phase is significant due to its spatial isotropic symmetry.  In spite of the pristine nature 
this phase no longer preserves two band model symmetry. On the other hand,  $d$-wave phases 
are hardly affected by doping and their superconductivity is more persistent against 
perturbation. The chirality or non-chirality of Cooper pairs
 in these phases is undetermined, however.
As shown in  Fig.~\ref{figure:g0-Tc}(b), at low temperature $\alpha^{+}_{s}\approx 0.6$  
for $\Phi^{+}_{s}$, and $\alpha_{d_{x^2-y^2}}^{+}=\alpha_{d_{xy}}^{+}\equiv\alpha^{+}_{d}$  
is approximately equal to unity and varies little with temperature.

\begin{figure}[ht]
\centering
\includegraphics[width=7cm]{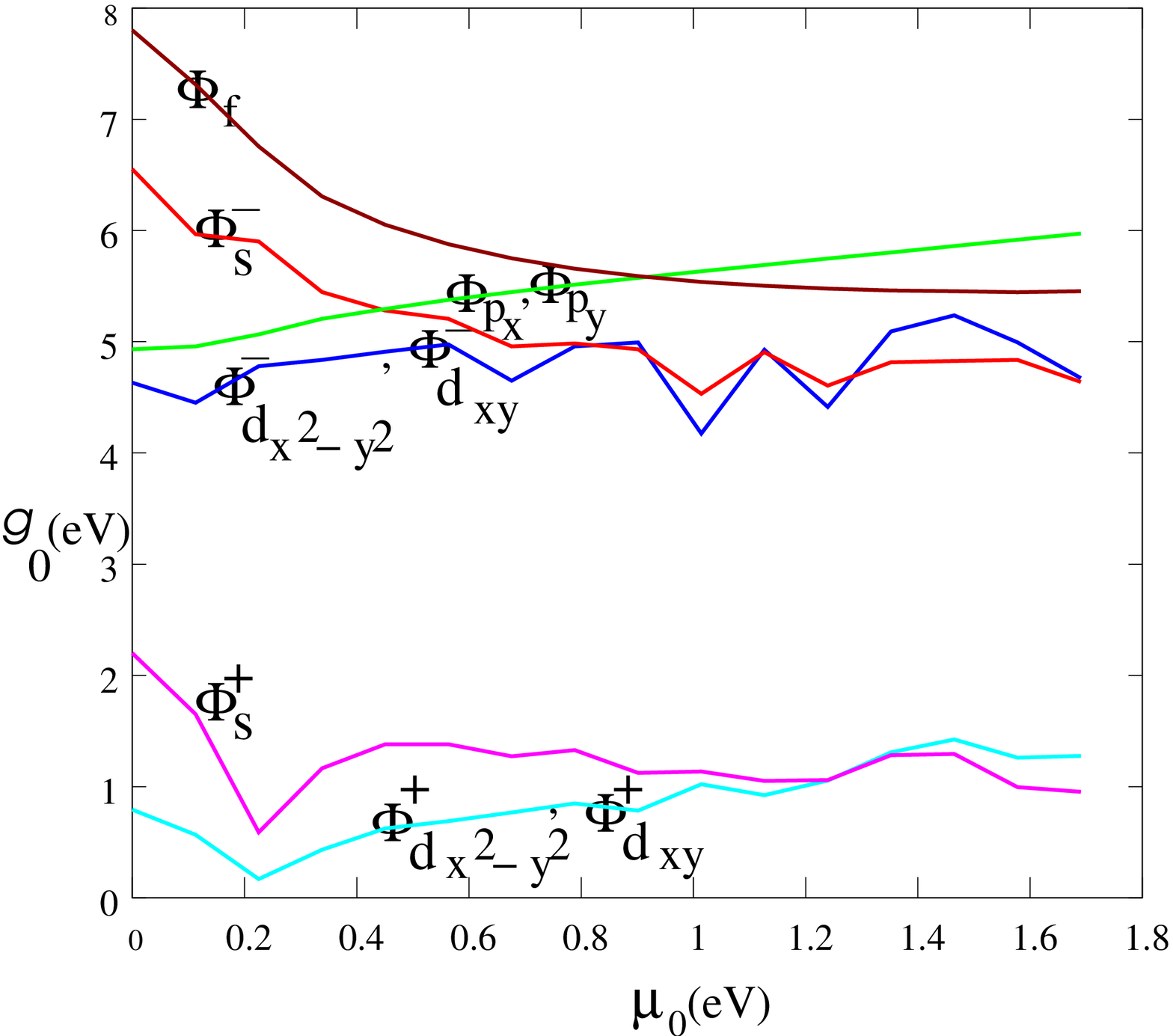}
\caption{(Color online) This diagram illustrates interaction potential $g_0$ in terms of chemical potential $\mu_0$ at $T_c=0.1$ K. Upon electron doping to a critical chemical potential
$\mu_{o-v}=0.22 eV$ (van Hove singularity) for symmetries $\Phi^{+}_{d_{x^2-y^2}}$, $\Phi^{+}_{d_{xy}}$,
  and $\Phi^{+}_{s}$  the pairing potential decreases, then increases until a second critical value
$\mu_{o-c}$=1.3 eV at which a phase transition to  $\Phi^{+}_{s}$ occurs.}
 \label{figure:umu} 
\end{figure}

At a given critical temperature $T_c$ and chemical potential $\mu_0$, for each of nine possible 
superconducting phases,  Eqs.~\ref{eq:gap-Equation-text}, 
\ref{eq:gapequation-matrix-unhermit-elements-text},
 and \ref{eq:quad-eigenvalue-text} were evaluated numerically over the BZ of Li$C_6$  to find
the corresponding pairing potential $g_0=\frac{1}{J_{sy}}$ and  ${\alpha_{sy}}$ coefficient.
Smaller $g_{0}$ means less  Cooper pair formation energy is required. 
Fig.~\ref{figure:g0-Tc}(a) provides the phase boundaries for $T_c$  in terms of the 
pairing  potential $g_0$ for Li$C_6$ in which $\mu_0=0$. 
 For a given transition temperature $T_c$, by changing the chemical potential $\mu_0$  
of Li$C_6$  via  gating,
 one can engineer the pairing potential $g_0$. Figure~\ref{figure:umu} gives a $g_0$-$\mu_0$ phase 
boundary diagram at $T_c=0.1$ K. As illustrated in this figure, similarly to pristine graphene,  
decoration with Li atoms makes it is possible to change the
dominant pairing and to have a symmetry-change phase transition from $d$ to `` distorted $s$-wave.'' 
Changing $\mu_{o}$ up to $\mu_{o-v}\approx 0.22 eV$ so that the distance between the Fermi 
energy and the saddle points decreases, leads to a decrease in $g_{0}$. Continuously increasing 
$\mu_{o}$ up to 0.5 eV  causes $g_{0}$ to increase for both $d$-wave 
and  ``distorted $s$-wave''  pairing, and after that a smooth decrease proceeds. 
For both symmetries at 
critical $\mu_{o-c}=1.3$ eV mixed state exist. 

 Up to $\mu_{o-c}=1.3$ eV, the flat band plays a primary role in formation of 
Cooper pairs with lowest energy. The Bloch wave function of this band consists 
of $d$ and $p$ character, therefore  $\Gamma_{12}$,  $\Gamma_{15}$,  $\Gamma_{45}$ 
and  $\Gamma_{48}$ in Eq.~\ref{eq:gapequation-matrix-unhermit-elements-text} carry 
minus signs. This makes it evident from Eq.~\ref{eq:quad-eigenvalue-text} that 
$d-$wave pairing is dominant. 
 Beyond that, the uneven part of the ``flat band'' and also upper bands 
assume a major role. These bands consist of $d$, $p$, $s$, and $f$ character 
Bloch wave functions (as defined in earlier sections) with a significantly 
low density of states. In this case  $\Gamma_{12}$,  $\Gamma_{15}$,  
$\Gamma_{45}$ and  $\Gamma_{48}$ change their sign, hence $s$-wave pairing 
is favored.

 Numerically we have demonstrated that electron pairing $g_0$ in the limit of pristine graphene 
is minimal for all dopings. Our calculations indicate that any perturbation 
 of the flat band  reduces T$_c$. The flat band can
be perturbed through  electron hopping from decorating atoms to carbon sites ($t_{1}^{LiC}$) or 
by hopping symmetry breaking index $w_t$. For fixed doping at $\bar{n}=0.11$ electron per carbon site 
 and for fixed $w_t=0.94$ as obtained for lithium decorated, in a variety of Li-C hopping between 
 0.3-0.4eV,  
numerical calculation doesn't show significant altering of pair interaction
 potential $g_0$  in $s$- and $d$-wave phases. 
But , as one could see there is not an explicit behavior in a general coupling strength. 
A result is that a general aspect of superconducting pairing in LiC$_6$ 
and pristine graphene
is almost the same in the $d_{x^2-y^2}$ and $d_{xy}$ phases due to
robustness of the flat band against perturbation.

To summarize, our calculations indicate that  $d$-wave phases 
 exist and are dominant symmetry of pairing in both pristine and Li decorated graphene. Pure $s$-wave
phase does not appear in LiC$_6$, and $s$-wave superconductivity  in metal decorated graphene 
is disfavored because of spatially increased overlap for $s$-wave symmetry. 
These  results show that while degree of doping  plays a major role in the graphene 
superconductivity, perturbation effects of decorating atoms finally determine the phase
diagram.
Our work also provides a new type of classification of superconducting phases 
in Li$C_6$-like nanostructures,
and certain aspects of the formalism may be useful in modeling the recently observed superconductivity
in magic angle bilayer graphene.\cite{Cao2018} 


\section*{Acknowledgments}
R. Gholami acknowledges support that allowed an extended
visit to the University of California Davis during part of this work.
W.E.P. was supported by NSF grant DMR-1207622.
\section*{Author contributions}
R. M. proposed the idea. R. G. and R. M. did the analytical derivation. R. G.  carried out the numerical calculations under supervision of R. M. and W. E. P. and  S. M.   performed the DFT calculations.  All authors analyzed the results and wrote the article.
\section*{Additional information}
\textbf{ Competing interests:} The authors declare no competing interests.
\section*{Appendices}
\appendix
\section{ Accurate Tight Binding Model for Lithium decorated Graphene  }
The Hamiltonian of non-interacting LiC$_6$ is 
\begin{equation}
 \hat{H}_{N}=-\sum_{i\alpha}\sum_{j \beta,\sigma } t_{i\alpha,j\beta}^{\sigma,\sigma}c_{i\alpha\sigma}^{\dagger}c_{j\beta\sigma} +\sum_{i\alpha ,\sigma }(\epsilon_{i\alpha}-\mu_{o} ) \hat{n}_{i\alpha\sigma}.
\label{eq:H-normal-app}
\end{equation}
Eq~\ref{eq:H-normal-app} can be diagonalized in terms of Bloch 
eigenfunctions of the form
\begin{eqnarray}
\left| \Psi_{\vec k} \right\rangle  = \frac{1}{\sqrt{N}}\sum\limits_{n = 1}^N {\sum\limits_{\alpha = 1}^7 {\mathscr{C}_{{\alpha }}}} e^{i\vec k.{{\vec r}_{n{\alpha}}}}\left| \phi_{n\alpha} \right\rangle 
\label{eq:bloch-state-0}
\end{eqnarray}
 in which ${\vec r}_{n\alpha}={\vec r}_{n}+{\vec d}_{\alpha}$ and ${\vec r}_{n}$ 
is $n$th Bravais lattice site vector position and ${\vec d}_{\alpha}$ is vector 
position of the $\alpha$-th subsite with respect to unit cell $n$. 
$\left| \phi_{n\alpha} \right\rangle=
   \left| \phi_{n\alpha}({\vec r}-{\vec r}_{n}-{\vec d}_{\alpha}) \right\rangle$ 
is the atomic $\pi$ electron ket of atom $\alpha$ of cell $n$. 
By inserting Eqs.~\ref{eq:H-normal-app}
 and \ref{eq:bloch-state-0}
 into $\hat{H}_{N}(\vec k)\left| \Psi_{\vec k}\right\rangle=E({\vec k})\left|
          \Psi_{\vec k}\right\rangle,$  
 the equation 
for the coefficients becomes
\begin{eqnarray}
\sum^{7}_{\beta=1} \varepsilon_{\alpha\beta}({\vec k}) \mathscr{C}_{\beta }+(\epsilon_{\alpha}-\mu_{o})\mathscr{C}_{\alpha } = E({\vec k})\mathscr{C}_{\alpha } \;\;\mbox{where}\;\; \varepsilon_{\alpha\beta}({\vec k})=-\frac{1}{N}\sum_{ij}t^{\sigma\sigma}_{i\alpha j\beta} e^{i{\vec k}.({\vec r}_{i\alpha}-{\vec r}_{j\beta})}.
\label{eq:schro-bloch-state-1}
\end{eqnarray}
Eq.~\ref{eq:schro-bloch-state-1} can be expressed in following matrix form
\begin{eqnarray}
\left( {\begin{array}{*{20}{c}}
{{\varepsilon_{LiLi}}(\vec k) + {\epsilon _{Li}} - {\mu _{o}}}&\vline& {{h_{LiA}}(\vec k)}&{{h_{LiB}}(\vec k)}\\
\hline
{h_{LiA}^\dag (\vec k)}&\vline& {{h_{AA}}(\vec k) + {\epsilon _A} - {\mu _{o}}}&{{h_{AB}}(\vec k)}\\
{h_{LiB}^\dag (\vec k)}&\vline& {{h_{BA}}(\vec k)}&{{h_{BB}}(\vec k) + {\epsilon _B} - {\mu _{o}}}
\end{array}} \right)\left( {\begin{array}{*{20}{c}}
{{{\mathscr{C}}_{Li}}({E_i}(\vec k))}\\
\hline
{{\xi _A}({E_i}(\vec k))}\\
{{\xi _B}({E_i}(\vec k))}
\end{array}} \right) = {E_i}(\vec k)\left( {\begin{array}{*{20}{c}}
{{{\mathscr{C}}_{Li}}({E_i}(\vec k))}\\
\hline
{{\xi _A}({E_i}(\vec k))}\\
{{\xi _B}({E_i}(\vec k))}
\end{array}} \right)
\label{eq:k-schr-1}
\end{eqnarray}
where
\begin{eqnarray}
\varepsilon_{LiLi}(\vec{k}) &=& 2{t}_1^{LiLi}\left( {\cos \vec k.{{\vec \xi }_1} + \cos \vec k.{{\vec \xi }_2} + \cos \vec k.{{\vec \xi }_3}} \right) + 2{t}_2^{LiLi}\left( {\cos \vec k.({{\vec \xi }_1} - {{\vec \xi }_2}) + \cos  \vec k.({{\vec \xi }_1} - {{\vec \xi }_3}) + \cos \vec k.({{\vec \xi }_2} - {{\vec \xi }_3})} \right) \nonumber\\
&+& 2{t}_3^{LiLi}\left( {\cos 2\vec k.{{\vec \xi }_1} + \cos 2\vec k.{{\vec \xi }_2} + \cos 2\vec k.{{\vec \xi }_3}} \right) +  \ldots 
\label{eq:block-diag-Li}
\end{eqnarray}
In Eq.~\ref{eq:k-schr-1}, $h$-sub-block matrices are C-C or Li-C dispersion matrices and we have $h_{AA}=h_{BB}^{*}$, $h_{AB}=h_{BA}^{\dagger}$.
The carbon-carbon dispersion matrices i.e.  $\varepsilon _{{(A,B)}_i{(A,B)}_j}(\vec k)$ are
\begin{eqnarray}
{h_{AA}}(\vec k) 
 = \left( {\begin{array}{*{20}{c}}
{\alpha(\vec k)}&{\beta(\vec k)}&{\gamma(\vec k)}\\
{\beta^{*}(\vec k)}&{\alpha(\vec k)}&{\theta(\vec k)}\\
{\gamma^{*}(\vec k)}&{\theta^{*}(\vec k)}&{\alpha(\vec k)}
\end{array}} \right),~~ 
{h_{AB}}(\vec k) =
  \left( {\begin{array}{*{20}{c}}
{{\tau _1}(\vec k)}&{{d_2}(\vec k)}&{{d_3}(\vec k)}\\
{{d_2}(\vec k)}&{{\tau _3}(\vec k)}&{{d_1}(\vec k)}\\
{{d_3}(\vec k)}&{{d_1}(\vec k)}&{{\tau _2}(\vec k)}.
\end{array}} \right)
\label{eq:block-diag-shrunk}
\end{eqnarray}
  The Li-C dispersion row matrices i.e. $\varepsilon _{Li{A_i}}(\vec k)$  and 
   $\varepsilon _{Li{B_i}}(\vec k)$ are
\begin{eqnarray}
{h_{LiA}}(\vec k) =[d_{1c}(\vec{k})~~d_{3c}(\vec{k})~~d_{2c}(\vec{k})]=
 - {t_{1}^{LiC}}{e^{i k_z h}}\left( {\begin{array}{*{20}{c}}
{{e^{i\vec k.{{\vec \delta }_1}}}}&{{e^{i\vec k.{{\vec \delta }_3}}}}&{{e^{i\vec k.{{\vec \delta }_2}}}}
\end{array}} \right),~~~
{h_{LiB}}(\vec k) =  {e^{i k_z h}}{h_{LiA}}^{*}(\vec k) 
\label{eq:block-offdiag-Li}
\end{eqnarray}
where $e^{k_zh}$ factor takes 1 by confinement. 
New variables, $\vec{k}$ dependent on-site energy and chemical potential has been defined as  
\begin{eqnarray}
\varepsilon_0({\vec{k}})&=& \epsilon_{Li}- \mu_{o}+\varepsilon_{LiLi}({\vec k}) \nonumber\\
\varepsilon_1({\vec{k}})&=&  \epsilon_{A}- \mu_{o}+\varepsilon_{A_1A_1}({\vec k})\nonumber\\
\varepsilon_2({\vec{k}})&=&  \epsilon_{B}- \mu_{o}+\varepsilon_{B_1B_1}({\vec k})
\label{eq:k-dep-chim-def}
\end{eqnarray}
Shorthand notation has been introduced as follows,
\begin{eqnarray}
\alpha (\vec{k})\equiv{\varepsilon _{{A_i}{A_i}}}(\vec k) &= &{\varepsilon _{{B_{i}}{B_{i}}}}(\vec k)=
-t_0- 2{t_5}\left( {\cos \vec k.{{\vec \xi }_1} + \cos \vec k.{{\vec \xi }_2} + \cos \vec k.{{\vec \xi }_3}} \right)\nonumber\\
 \beta (\vec{k})\equiv{\varepsilon _{{A_1}{A_2}}}(\vec k) &=&\varepsilon _{{A_2}{A_1}}^*(\vec k) = -{t_2}{e^{i\vec k.({{\vec \delta }_3} - {{\vec \delta }_1})}}\left[ {1 + w_t\left( {{e^{ - i\vec k.{{\vec \xi }_3}}} + {e^{i\vec k.{{\vec \xi }_1}}}} \right)} \right]\nonumber\\
\gamma (\vec{k})\equiv{\varepsilon _{{A_1}{A_3}}}(\vec k) &=& \varepsilon _{{A_3}{A_1}}^*(\vec k)= -{t_2}{e^{i\vec k.({{\vec \delta }_2} - {{\vec \delta }_1})}}\left[ {1 + w_t\left( {{e^{ - i\vec k.{{\vec \xi }_2}}} + {e^{i\vec k.{{\vec \xi }_1}}}} \right)} \right]\nonumber\\
\theta(\vec{k})\equiv{\varepsilon _{{A_2}{A_3}}}(\vec k)& = &\varepsilon _{{A_3}{A_2}}^*(\vec k) = -{t_2}{e^{i\vec k.({{\vec \delta }_2} - {{\vec \delta }_3})}}\left[ {1 + w_t\left( {{e^{ - i\vec k.{{\vec \xi }_2}}} + {e^{i\vec k.{{\vec \xi }_3}}}} \right)} \right]
\label{eq:epAA}
\end{eqnarray}
in which $w_t=\frac{ t^{'}_{1}}{t_{1}}=\frac{ t^{'}_{2}}{t_{2}}$ and $\vec{\xi_{i}}=\vec{\tau_i}+2\vec{\delta_i}$ the $d$ and $\tau$,functions are given by
\begin{eqnarray}
{\tau _1}(\vec k) = -t_{1}^{'}{e^{i\vec k.{{\vec \tau }_1}}}\left[ {1 + \frac{{{t_3}}}{{{t^{'}_{1}}}}{e^{ - i\vec k.{{\vec \xi }_1}}} + \frac{{{t_4}}}{{{t^{'}_{1}}}}\left( {{e^{i\vec k.{{\vec \xi }_2}}} + {e^{i\vec k.{{\vec \xi }_3}}}} \right)} \right],~ ~{d_1}(\vec k) = -t_1{e^{i\vec k.{{\vec \delta }_1}}}\left[ {1 + \frac{{{t_3}}}{{{t_1}}}{e^{ - i\vec k.{{\vec \xi }_1}}} + \frac{{{t_4}}}{{{t_1}}}\left( {{e^{i\vec k.{{\vec \xi }_2}}} + {e^{i\vec k.{{\vec \xi }_3}}}} \right)} \right]\nonumber \\
{\tau _2}(\vec k) = -t_{1}^{'} {e^{i\vec k.{{\vec \tau }_2}}}\left[ {1 + \frac{{{t_3}}}{{{t^{'}_{1}}}}{e^{ - i\vec k.{{\vec \xi }_2}}} + \frac{{{t_4}}}{{{t^{'}_{1}}}}\left( {{e^{i\vec k.{{\vec \xi }_3}}} + {e^{i\vec k.{{\vec \xi }_1}}}} \right)} \right],~ ~ {d_2}(\vec k) = -t_1{e^{i\vec k.{{\vec \delta }_2}}}\left[ {1 + \frac{{{t_3}}}{{{t_1}}}{e^{ - i\vec k.{{\vec \xi }_2}}} + \frac{{{t_4}}}{{{t_1}}}\left( {{e^{i\vec k.{{\vec \xi }_3}}} + {e^{i\vec k.{{\vec \xi }_1}}}} \right)} \right]\nonumber \\
{\tau _3}(\vec k) =  -t_{1}^{'}{e^{i\vec k.{{\vec \tau }_3}}}\left[ {1 + \frac{{{t_3}}}{{{t^{'}_{1}}}}{e^{ - i\vec k.{{\vec \xi }_3}}} + \frac{{{t_4}}}{{{t^{'}_{1}}}}\left( {{e^{i\vec k.{{\vec \xi }_1}}} + {e^{i\vec k.{{\vec \xi }_2}}}} \right)} \right],~~ {d_3}(\vec k) = -t_1{e^{i\vec k.{{\vec \delta }_3}}}\left[ {1 + \frac{{{t_3}}}{{{t_1}}}{e^{ - i\vec k.{{\vec \xi }_3}}} + \frac{{{t_4}}}{{{t_1}}}\left( {{e^{i\vec k.{{\vec \xi }_1}}} + {e^{i\vec k.{{\vec \xi }_2}}}} \right)} \right].
\label{eq:epAB}
\end{eqnarray}
Also
\begin{eqnarray}
{\xi _A}(\vec k) = \left( {\begin{array}{*{20}{c}}
{{{\mathscr C}_{{A_1}}}({E_i}(\vec k))}\\
{{{\mathscr C}_{{A_2}}}({E_i}(\vec k))}\\
{{{\mathscr C}_{{A_3}}}({E_i}(\vec k))}
\end{array}} \right)\quad ; \quad {\xi _B}(\vec k) = \left( {\begin{array}{*{20}{c}}
{{{\mathscr C}_{{B_1}}}({E_i}(\vec k))}\\
{{{\mathscr C}_{{B_2}}}({E_i}(\vec k))}\\
{{{\mathscr C}_{{B_3}}}({E_i}(\vec k))}
\end{array}} \right).
\label{eq:shrunk-eigenstates}
\end{eqnarray}
 Since the Li is an inversion center of the two-dimension tight binding model, 
the Bloch wave function should respect inversion symmetry. Using $\Psi_{\vec k}(\vec r)
= \langle \vec r|\Psi_{\vec k}\rangle,$ the condition is
\begin{eqnarray}
\Psi_{-\vec k}(-\vec r) = Ce^{i\phi}  \Psi_{\vec k} (\vec r),   
\label{eq:bloch-inversion} 
\end{eqnarray}
where $C$ is $\pm1$ when all subsites in hexagons are carbon, that is 
$\epsilon_{A_{i}}=\epsilon_{B_{i}}$. This condition is satisfied if 
${\mathscr C}_{A_{i}}$ is proportional to ${\mathscr C}^{*}_{B_{i}}$
\begin{eqnarray}
{\mathscr C}_{A_{m}}({E_i}(\vec k))=f_{i}(\vec k){\mathscr C}^{*}_{B_{m}}({E_i}(\vec k)),\;\;\mbox{ and}\;\;{\mathscr C}_{Li}({E_i}(\vec k))=f_{i}(\vec k){\mathscr C}^{*}_{Li}({E_i}(\vec k)), \;\;\;i=1,2,...,7
\label{eq:invariant-cond} 
\end{eqnarray}
where $f_{i}(\vec k)$ is a coefficient to be determined. By inserting ${\mathscr C}_{A_{i}}=|{\mathscr C}_{A_{i}}|e^{i\phi_{A_{i}}}$ and  ${\mathscr C}_{B_{i}}=|{\mathscr C}_{B_{i}}|e^{i\phi_{B_{i}}}$ into Eq.~\ref{eq:invariant-cond} we have
\begin{eqnarray}
f_{i}(\vec k)=\frac{|{\mathscr C}_{A_{m}}({E_i}(\vec k))|}
     {|{\mathscr C}_{B_{m}}({E_i}(\vec k))|} 
     e^{i(\phi_{A_{i}}+\phi_{B_{i}})}= C_{i}e^{i\phi_{i}}, \;\;\;i=1,2, ...,7.
\label{eq:frac-Ca-Cb} 
\end{eqnarray}
In the next subsection we use Eqs.~\ref{eq:invariant-cond} and \ref{eq:frac-Ca-Cb} to reduce 
the eigenvalue Eq.~\ref{eq:schro-bloch-state-1} in matrix form to $3\times 3$ to obtain 
uncoupled shrunken graphene band structure and Bloch wave function coefficients.  
\section{ Uncoupled $C_6$ Dispersion Relations }
By first neglecting the lithium-carbon hopping $t_{1}^{LiC}\rightarrow 0$, 
 the shrunken graphene Hamiltonian Eq.~\ref{eq:k-schr-1} can be diagonalized 
exactly. Our notation is 
  \begin{eqnarray}
 \left( {\begin{array}{*{20}{c}}
{{\varepsilon _0}(\vec k)}&\vline& 0&0&0&\vline& 0&0&0\\
\hline
0&\vline& {{\varepsilon _1}(\vec k)}&{ \beta (\vec k)}&{ \gamma (\vec k)}&\vline& {{\tau_1}(\vec k)}&{{d_2(\vec k)}}&{{d_3(\vec k)}}\\
0&\vline& {  {\beta ^*}(\vec k)}&{{\varepsilon _1}(\vec k)}&{  \theta (\vec k)}&\vline& {{d_2(\vec k)}}&{{\tau_3}(\vec k)}&{{d_1(\vec k)}}\\
0&\vline& {  {\gamma ^*}(\vec k)}&{  {\theta ^*}(\vec k)}&{{\varepsilon _1}(\vec k)}&\vline& {{d_3(\vec k)}}&{{d_1(\vec k)}}&{{\tau_2}(\vec k)}\\
\hline
0&\vline& {\tau_1^*(\vec k)}&{d_2^{*}(\vec k)}&{d_3^*(\vec k)}&\vline& {{\varepsilon _2}(\vec k)}&{  {\beta ^*}(\vec k)}&{  {\gamma ^*}(\vec k)}\\
0&\vline& {d_2^{*}(\vec k)}&{\tau_3^*(\vec k)}&{d_1^{*}(\vec k)}&\vline& {  \beta (\vec k)}&{{\varepsilon _2}(\vec k)}&{  {\theta ^*}(\vec k)}\\
0&\vline& {d_3^*(\vec k)}&{d_1^{*}(\vec k)}&{\tau_2^*(\vec k)}&\vline& {  \gamma (\vec k)}&{  \theta (\vec k)}&{{\varepsilon _2}(\vec k)}
\end{array}} \right)\left( {\begin{array}{*{20}{c}}
{{C_{Li}}({E^{0}_i}(\vec k))}\\
\hline
{{C_{A_{1}}}({E^{0}_i(\vec k)})}\\
{{C_{A_{2}}}({E^{0}_i}(\vec k))}\\
{{C_{A_{3}}}({E^{0}_i}(\vec k))}\\
{{C_{B_{1}}}({E^{0}_i}(\vec k))}\\
{{C_{B_{2}}}({E^{0}_i}(\vec k))}\\
{{C_{B_{3}}}({E^{0}_i}(\vec k))}
\end{array}} \right)={E^{0}_i(\vec k)} \left( {\begin{array}{*{20}{c}}
{{C_{Li}}({E^{0}_i}(\vec k))}\\
\hline
{{C_{A_{1}}}({E^{0}_i(\vec k)})}\\
{{C_{A_{2}}}({E^{0}_i}(\vec k))}\\
{{C_{A_{3}}}({E^{0}_i}(\vec k))}\\
{{C_{B_{1}}}({E^{0}_i}(\vec k))}\\
{{C_{B_{2}}}({E^{0}_i}(\vec k))}\\
{{C_{B_{3}}}({E^{0}_i}(\vec k))}
\end{array}} \right)
\label{eq:non-perturb-sch}
\end{eqnarray}
  The non trivial eigenvalues of uncoupled Hamiltonian Eq.~\ref{eq:non-perturb-sch} are given by 
\begin{eqnarray}
{E_{sh;n}}(\vec k) ={E_{sh,ml}}(t_{i},\vec{\xi_{i}},\vec{k}) =- {\mu _{o}} +\alpha (\vec k)  +u_{m}\Pi_{0}(\vec k)+ u^ *_{m} \Pi^{*}_{0}(\vec k) + \frac{1}{2}\left[ {{\varepsilon _A} + {\varepsilon _B} + {{( - 1)}^l}\sqrt {{{\left( {{\varepsilon _A} - {\varepsilon _B}} \right)}^2} + 4{w_{m}}(\vec k)}   } \right]
\label{eq:band-structure of shrunk graphene}
\end{eqnarray}
  where $n$ is band index defined as
\begin{eqnarray}
n = (2l + 1) + {( - 1)^l}m;\qquad 
m=     \left\{ \begin{array}{ccl}
      0,\; 1\;,\;\; 2& \mbox{for l=0}
      & \mbox{conduction}\\ -1,-2,-3 & \mbox{for l=1} & \mbox{valence} 
             \end{array}\right.
\label{eq:band index} 
\end{eqnarray}
Here $E_{sh,1}$, $E_{sh,2}$  and $E_{sh,3}$ are conduction bands which corresponds to $l=0$ and
 $ m=0,1,2$ and $E_{sh,4}$, $E_{sh,5}$  and $E_{sh,6}$ are valence  bands which correspond to $l=1, \; m=-1,-2,-3$.  

At the $\Gamma$ point, when $\varepsilon_A=\varepsilon_B=\varepsilon^{c}$ , the shrunk graphene eigenstates  $\left| {{\phi_{n}}(0)} \right\rangle =(C_{A_1}~C_{A_2}~C_{A_3}~C_{B_1}~C_{B_2}~C_{B_3})^T$ over the hexagonal subsites take the forms similar to conventional $s$, $d$ and $p$ orbitals,
\begin{eqnarray}
\left|f\right\rangle&=&(1~~1~~1-1-1-1)^{T}, ~~~
\left|p_{x}\right\rangle=(1-1~~0-1~~1~~0)^{T}, ~~~
\left|p_{y}\right\rangle=(1~~1-2-1-1~~2)^{T}, \nonumber \\
\left|d_{x^2-y^2}\right\rangle&=&(1~~1-2~~~1~~1-2)^{T}, ~~~
\left|d_{xy}\right\rangle=(1-1~~0~~1-1~~0)^{T},~~~
\left|S\right\rangle=(1~~1~~1~~1~~1~~1)^{T}, 
\label{eq:gamma-energy-shrunk}
\end{eqnarray}
 with energies,
 \begin{eqnarray}
E_f&=& E^{+}_{\gamma}(0)=\mu_s+[(t_{1}^{'}+2t_1)+3t_3+6t_4]\nonumber \\
E_s&=& E^{-}_{\gamma}(0)=\mu_s-[(t_{1}^{'}+2t_1)+3t_3+6t_4]\nonumber \\
E_p&=&E^{+}_{\alpha}(0)=E^{+}_{\beta}(0)=\mu_d+(t_{1}^{'}-t_1)\nonumber \\
E_d&=&E^{-}_{\alpha}(0)=E^{-}_{\beta}(0)=\mu_d-(t_{1}^{'}-t_1)
 \end{eqnarray}
 where $\mu_s=\varepsilon^{c}-\mu_0-2(t_2+2t^{'}_{2})-6t_5$ and $\mu_d=\varepsilon^{c}-\mu_0+(t_2+2t^{'}_{2})-6t_5.$
For general $\vec k$ the uncoupled shrunken graphene eigenfunction Eq.~\ref{eq:shrunk-eigenstates-0} can be written in terms of  $\left|S\right\rangle$, $\left|f\right\rangle$, $\left|p_x\right\rangle$,$\left|p_y\right\rangle$, $\left|d_{xy}\right\rangle$ and $\left|d_{x^2-y^2}\right\rangle$ as
\begin{eqnarray}
\left| {{\phi_{n}}(\vec k)} \right\rangle  = \left(f^{n}_{s}(\vec k)\left|S\right\rangle+ f^{n}_{f}(\vec k)\left|f\right\rangle\right)+ \left(f^{n}_{p_y}(\vec k)\left|p_y\right\rangle+if^{n}_{d_{xy}}(\vec k)\left|d_{xy}\right\rangle\right)+\left(f^{n}_{d_{x^2-y^2}}(\vec k)\left|d_{x^2-y^2}\right\rangle+if^{n}_{p_{x}}(\vec k)\left|p_{x}\right\rangle\right).
\label{eq:linear-com}
\end{eqnarray}
In the particular case of pristine graphene in which $w_t=1$ and $\vec{\tau_1}=\vec{\delta_1} ,\vec{\tau_2}=\vec{\delta_2}, \vec{\tau_3}=\vec{\delta_3}$, 
hence $\beta=\theta=\gamma^{*}$ and also $\varepsilon_A=\varepsilon_B$ so 
$C=\pm 1$. Therefore eigenvectors  take following form
\begin{eqnarray}
\left|  \varphi _{m,l}^0(\vec k) \right\rangle= \frac{1}{{\sqrt 6 }}\left(
{{\omega _m}}~~
{{\omega _m}^*}~~
1~~
{ (-1)^{l} \frac{{\eta _m^ * }}{{\left| {{\eta _m}} \right|}}{\omega _m}^*}~~
{ (-1)^{l} \frac{{\eta _m^ * }}{{\left| {{\eta _m}} \right|}}{\omega _m}\:}~~
{ (-1)^{l} \frac{{\eta _m^ * }}{{\left| {{\eta _m}} \right|}}}
 \right)^T, ~~  \omega_m  = {e^{i2\pi m/3}}, 
\label{eq:pristine-eigenvectors}
\end{eqnarray}
where  $ m=1,2 ,3; \;\;l=0, 1$ and $
\eta_{m}(\vec k)=d_{2}(\vec k)+\omega_{m}d_{1}(\vec k)+\omega _{m}^{*}d_{3}(\vec k)$.
The eigenvalues are
\begin{eqnarray}
E^{0}_{m,l}=\varepsilon_{A_1A_1}(\vec k)+\omega_{m}\beta(\vec k)+\omega_{m}^{*}\beta^{*}(\vec k)+(-1)^{l}t_{1}|\eta_{m}(\vec k)|
\label{eq:pristine-eigenvalues}
\end{eqnarray}
By comparing pristine eigenvectors Eq.~\ref{eq:pristine-eigenvectors} with general shrunken graphene eigenvectors Eq.~\ref{eq:linear-com} it is found that for $m=1$, $f^{1}_{d_{x^2-y^2}}(\vec k)=-f^{1}_{p_x}(\vec k)=\frac{(1-e^{i\phi_1})}{2\sqrt{2}}$ which corresponds to $d_{x^2-y^2}-ip_{x}$, 
also  
$f^{1}_{p_y}(\vec k)=-f^{1}_{d_{xy}}(\vec k)=\frac{(1+e^{i\phi_1})}{2\sqrt{2}}$ 
which corresponds to $p_y-id_{xy}$, and the coefficients of $s$ and $f$ are zero. 
For $m=2$, $f^{2}_{d_{x^2-y^2}}(\vec k)=f^{2}_{p_x}(\vec k)=\frac{(1-e^{i\phi_2})}{2\sqrt{2}}$ 
which corresponds to 
$d_{x^2-y^2}+ip_{x}$ also  
$f^{2}_{p_y}(\vec k)=f^{2}_{d_{xy}}(\vec k)=\frac{(1+e^{i\phi_2})}{2\sqrt{2}}$ 
which corresponds to $p_y+id_{xy}$ while $s$ and $f$ coefficients are zero. 
For $m=3$, $f^{3}_{s}(\vec k)=\frac{(1-e^{i\phi_3})}{2}$ and 
$f^{3}_{f}(\vec k)=\frac{(1+e^{i\phi_3})}{2}$ and other coefficients are zero. 
Here, $e^{i\phi_m}=\frac{\eta_{m}^{*}}{|\eta_m|}$.

The Hamiltonian $\hat H_{N}^{shr}$  for the broken symmetry (shrunk
graphene) is $6 \times 6$ in terms of $3\times 3$ subblocks
 \begin{eqnarray}
 \left( {\begin{array}{*{20}{c}}
{{h_{AA}}(\vec k)+\varepsilon_{A} - {\mu _{o}}}&{{h_{AB}}(\vec k)}\\
{{h_{BA}}(\vec k)}&{{h_{AA}^{*}}(\vec k)+\varepsilon_B - {\mu _{o}}}
\end{array}} \right)\left( {\begin{array}{*{20}{c}}
{{\xi _{A}^{0}}{(E_{sh;i}}(\vec k))}\\
{{\xi _{B}^{0}}(E_{sh;i}(\vec k))}
\end{array}} \right) = {E_{sh;i}}(\vec k)\left( {\begin{array}{*{20}{c}}
{{\xi _A^{0}}(E_{sh;i}(\vec k))}\\
{{\xi _B^{0}}(E_{sh;i}(\vec k))}
\end{array}} \right)
\label{eq:Schro-eq-shrunk}
 \end{eqnarray}
 To solve Schr\"odinger Eq.~\ref{eq:Schro-eq-shrunk}  we first separate the left hand side of 
Eq.~\ref{eq:Schro-eq-shrunk}  into two terms
 ${H_{sh}}(\vec k) = H_{sh}^0(\vec k) + H_{sh}^1(\vec k)$
 where
 \begin{eqnarray}
 H_{sh}^0(\vec k) =
 \left( {\begin{array}{*{20}{c}}
{{\varepsilon_1}(\vec{k}){I_{3 \times 3}}}&{{h_{AB}}(\vec k)}\\
{{h_{BA}}(\vec k)}&{{\varepsilon_2}(\vec{k}){I_{3 \times 3}}}
\end{array}} \right),~~
H_{sh}^1(\vec k) = \left( {\begin{array}{*{20}{c}}
{{{h'}_{aa}}(\vec k)}&{{0_{3 \times 3}}}\\
{{0_{3 \times 3}}}&{{{h'}_{bb}}(\vec k)}
\end{array}} \right) 
\label{eq:diag-shrunk-H}
\end{eqnarray}
 where ${h}_{aa}^{'}(\vec k)=({h}_{AA}(\vec k)-\alpha(\vec k)I_{3\times3})$.
In the nearest neighbor approximation it  is straightforward to show that 
$H_{sh}^{0}(\vec{k})$ and $H_{sh}^{1}(\vec{k})$ commute with each other, hence 
they have the same eigenvectors. In more generality one can consider approximately 
the eigenvectors of $H_{sh}^{1}$ to be the same as the eigenvectors of  
$H_{sh}^{0}$, therefore one obtains 
\begin{eqnarray}
\xi_{B}^{0}(\vec k)=Ce^{i\phi}\xi_{A}^{0*}: ~~~~
{E_{sh}}(\vec k) \approx  E_{sh}^0(\vec k) + E_{sh}^1(\vec k).
\label{eq:shrunk-eigenvalue-sepr}
\end{eqnarray}
with the first equation arising from similar 
conditions as for Eq.~\ref{eq:invariant-cond}. To find $E_{sh}(\vec k)$ we solve the 
following eigenvalue problems 
\begin{eqnarray}
\left( {\begin{array}{*{20}{c}}
{{{h'}_{aa}}(\vec k)}&{{0_{3 \times 3}}}\\
{{0_{3 \times 3}}}&{h_{aa}^{'*}}(\vec k)
\end{array}} \right)\left( {\begin{array}{*{20}{c}}
{\xi _A^0(\vec k)}\\
{\xi _B^0(\vec k)}
\end{array}} \right) = E_{sh}^1(\vec k)\left( {\begin{array}{*{20}{c}}
{\xi _A^0(\vec k)}\\
{\xi _B^0(\vec k)}
\end{array}} \right)
\label{eq:new-eigenvalues-problems-shrunk-diag}
\end{eqnarray}
and
\begin{eqnarray}
\left( {\begin{array}{*{20}{c}}
{ {\varepsilon_1(\vec k) } {I_{3 \times 3}}}&{{h_{AB}}(\vec k)}\\
{{h_{BA}}(\vec k)}&{{\varepsilon_2(\vec k) }} {I_{3 \times 3}}
\end{array}} \right)\left( {\begin{array}{*{20}{c}}
{\xi _A^0(\vec k)}\\
{\xi _B^0(\vec k)}
\end{array}} \right) = E_{sh}^0(\vec k)\left( {\begin{array}{*{20}{c}}
{\xi _A^0(\vec k)}\\
{\xi _B^0(\vec k)}
\end{array}} \right)
\label{eq:new-eigenvalues-problems-shrunk-offdiag}
\end{eqnarray}
 Eq.~\ref{eq:new-eigenvalues-problems-shrunk-diag} converts to following  eigenvalue problem
$
{{h'}_{aa}}(\vec k)\xi _A^0(\vec k) = E_{sh}^1(\vec k)\xi _A^0(\vec k)
$
and its complex conjugate with $A\leftrightarrow B$, defining 
\begin{eqnarray}
{c_0}(t_{2},\vec{\xi_{i}},\vec{k}) &= &{\beta}(\vec{k}) {\theta}(\vec{k}) {\gamma ^ * }(\vec{k}) +{ \gamma }(\vec{k}){\beta ^ * }(\vec{k}){\theta ^ * }(\vec{k}),~~
{c_1}(t_{2},\vec{\xi_{i}},\vec{k}) = {\left| \beta (\vec{k}) \right|^2} + {\left| \theta (\vec{k}) \right|^2} + {\left| \gamma (\vec{k}) \right|^2}\nonumber\\
\Pi_{0}(t_{2},\vec{\xi_{i}},\vec{k})  & =& {\left( {{\textstyle{{c_{0}(t_{2},\vec{\xi_{i}},\vec{k})} \over 2}} + i\sqrt {{{\left( {{\textstyle{{{c_1(t_{2},\vec{\xi_{i}},\vec{k})}} \over 3}}} \right)}^3} - {{\left( {{\textstyle{{{c_0(t_{2},\vec{\xi_{i}},\vec{k})}} \over 2}}} \right)}^2}} } \right)^{{\raise0.5ex\hbox{$\scriptstyle 1$}
\kern-0.1em/\kern-0.15em
\lower0.25ex\hbox{$\scriptstyle 3$}}}}.
\label{eq:c0-c1-def}
\end{eqnarray}
eigenvalue equation Eq.~\ref{eq:new-eigenvalues-problems-shrunk-diag} has the three different eigenvalues
\begin{eqnarray}
{E_{sh,m}^1(t_{2},\vec{\xi_{i}},\vec{k})} = u_{m}\Pi_{0}(t_{2},\vec{\xi_{i}},\vec{k})  + 
  u^ *_{m} {\Pi}^*_{0}(t_{2},\vec{\xi_{i}},\vec{k});\qquad u_{m}=\sqrt[3]{1} = {e^{4 i m\pi /3}};\;m=1,2,3
\label{eq:diag-eigenvalues}
\end{eqnarray}
note that ${E_{sh,m}^1(t_{2},\vec{\xi_{i}},\vec{k})}$ is function of second neighbor hopping $t_{<iA_{i},jA_{j}>}=t_{2}$ and $\vec{\xi_{i}}=\vec{\tau_{i}}+2 \vec{\delta_{i}}$ i.e. lattice bases vector and it does not depend on $\vec{\tau_{i}}$ and $\vec{\delta_{i}}$ separately. Now we calculate $E_{sh}^0(\vec k)$. Eq.~\ref{eq:new-eigenvalues-problems-shrunk-offdiag} can be separated into following equations
\begin{eqnarray}
{h_{AB}}(\vec k){\xi _{B}^{0}}(\vec k) = (E_{sh}^0(\vec k) - {\varepsilon _1}(\vec k)){\xi _{A}^{0}}(\vec k),~~
 {h_{BA}}(\vec k){\xi _{A}^{0}}(\vec k) = (E_{sh}^0(\vec k) - {\varepsilon _2}(\vec k)){\xi _{B}^{0}}(\vec k).
\label{eq:couple-eigen}
\end{eqnarray}
By multiplying first equation of Eq.~\ref{eq:couple-eigen} by $h_{BA}$ and second by $h_{AB}$ we have
\begin{eqnarray}
{h_{AB}}(\vec k){h_{BA}}(\vec k){\xi _{A}^{0}}(\vec k) = (E_{sh}^0(\vec k) - {\varepsilon _1}(\vec k))(E_{sh}^0(\vec k) - {\varepsilon _2}(\vec k)){\xi _{A}^{0}}(\vec k)\nonumber \\
{h_{BA}}(\vec k){h_{AB}}(\vec k){\xi _{B}^{0}}(\vec k) = (E_{sh}^0(\vec k) - {\varepsilon _1}(\vec k))(E_{sh}^0(\vec k) - {\varepsilon _2}(\vec k)){\xi _{B}^{0}}(\vec k).
\label{eq:new-eigen-prob2}
\end{eqnarray} 
Eq.~\ref{eq:new-eigen-prob2} is an eigenvalue problem where second equation is just complex conjugated of first one. By defining new matrix 
$G(\vec k) = {h_{AB}}(\vec k){h_{BA}}(\vec k) $
 and $w_i(\vec k) = [E_{sh,i}^0(\vec k)]^{2} -[ {\varepsilon _1}(\vec k)+{\varepsilon _2}(\vec k)]E_{sh,i}^0(\vec k)+{\varepsilon _1}(\vec k){\varepsilon _2}(\vec k)$, Eq.~\ref{eq:new-eigen-prob2} takes the form
\begin{eqnarray} 
G(\vec k){\xi _{A}^{0}}(E_{i}^{0}(\vec k)) = w_{i}(\vec k){\xi _{A}^{0}}(E_{i}^{0}(\vec k))
\label{eq:G-eigenvalue-prob}
\end{eqnarray} 
Schr\"odinger Eq.~\ref{eq:G-eigenvalue-prob} can be solved to find eigenvalues of  Eq.~\ref{eq:new-eigenvalues-problems-shrunk-offdiag} i.e. $E_{sh}^0(\vec k)$. Defining  
\begin{eqnarray}
{C_2} (t_{i},\vec{\xi_{i}},\vec{k}) &=& {G_{11}} + {G_{22}} + {G_{33}}\nonumber\\
{C_1}(t_{i},\vec{\xi_{i}},\vec{k})  &=& {\left| {{G_{12}}} \right|^2} + {\left| {{G_{13}}} \right|^2} + {\left| {{G_{23}}} \right|^2} - \left( {{G_{11}}{G_{22}} + {G_{11}}{G_{33}} + {G_{22}}{G_{33}}} \right)\nonumber\\
{C_0}(t_{i},\vec{\xi_{i}},\vec{k}) & =& {G_{13}}{({G_{12}}{G_{23}})^*} + G_{13}^*({G_{12}}{G_{23}}) - {G_{11}}{\left| {{G_{23}}} \right|^2} - {G_{22}}{\left| {{G_{13}}} \right|^2} - {G_{33}}{\left| {{G_{12}}} \right|^2} + {G_{11}}{G_{22}}{G_{33}}
\label{eq:Ci-coef-def}
\end{eqnarray}
where ${G_{ij}} = \sum\limits_{m = 1}^3 {{\varepsilon _{{A_i}{B_m}}}(\vec k){\varepsilon _{{B_m}{A_j}}}(\vec k)}$. Also introducing 
\begin{eqnarray}
\Pi_1(t_{i},\vec{\xi_{i}},\vec{k}) &= &{\left( {Q(t_{i},\vec{\xi_{i}},\vec{k})+ i\sqrt {{P(t_{i},\vec{\xi_{i}},\vec{k})^3} - {Q(t_{i},\vec{\xi_{i}},\vec{k})^2}} } \right)^{\frac{1}{3}}}\nonumber\\
Q(t_{i},\vec{\xi_{i}},\vec{k}) &= &\frac{{{C_0(t_{i},\vec{\xi_{i}},\vec{k})}}}{2} + \frac{{{C_1(t_{i},\vec{\xi_{i}},\vec{k})}{C_2(t_{i},\vec{\xi_{i}},\vec{k})}}}{6} + \frac{{C_2^3(t_{i},\vec{\xi_{i}},\vec{k})}}{{27}} \nonumber\\
P(t_{i},\vec{\xi_{i}},\vec{k})& = &\frac{{{C_1(t_{i},\vec{\xi_{i}},\vec{k})}}}{3} + \frac{{C_2^2(t_{i},\vec{\xi_{i}},\vec{k})}}{9}.
\label{eq:delta0-def}
\end{eqnarray}
where $t_{i}=t_{iAjB}$ are first, 3rd and 4th neighbor hopping integrals. Hence eigenvalues of  Eq.~\ref{eq:new-eigenvalues-problems-shrunk-offdiag} can be obtained, they are
\begin{eqnarray}
E_{sh;m,l}^0(t_{i},\vec{\xi_{i}},\vec{k}) = {\varepsilon _{{A_1}{A_1}}}(\vec k) - {\mu _{o}} + \frac{1}{2}\left[ {{\varepsilon _A} + {\varepsilon _B} + {{( - 1)}^l}\sqrt {{{\left( {{\varepsilon _A} - {\varepsilon _B}} \right)}^2} + 4{w_m}(t_{i},\vec{\xi_{i}},\vec{k})}   } \right] 
\label{eq:shrunk-eigen} 
\end{eqnarray}
where $l= 0, \;1$ and ${w_m}(t_{i},\vec{\xi_{i}},\vec{k})$ i.e. solutions of Schr\"odinger Eq.~\ref{eq:G-eigenvalue-prob} are
\begin{eqnarray}
w_{m}(t_{i},\vec{\xi_{i}},\vec{k}) = \frac{C_2(t_{i},\vec{\xi_{i}},\vec{k})}{3}+ u_{m}\Pi_1(t_{i},\vec{\xi_{i}},\vec{k})  + u^ *_{m} {\Pi}_{1}^{*}(t_{i},\vec{\xi_{i}},\vec{k})  \quad; u_{m} = \sqrt[3]{1} = {e^{4 i m\pi /3}}\quad;\;m=  1 , 2, 3.
\label{eq:cubic-w-solutions}
\end{eqnarray}
Hence from Eqs.~\ref{eq:shrunk-eigenvalue-sepr}, \ref{eq:diag-eigenvalues} and  \ref{eq:shrunk-eigen}     eigenvalues of Schr\"odinger Eq.~\ref{eq:Schro-eq-shrunk} can be obtained,
\begin{eqnarray}
{E_{sh;m,l}}(t_{i},\vec{\xi_{i}},\vec{k}) = E_{sh,m}^1(t_{i},\vec{\xi_{i}},\vec{k})+ {\varepsilon _{{A_1}{A_1}}}(\vec k) - {\mu _{o}} + \frac{1}{2}\left[ {{\varepsilon _A} + {\varepsilon _B} + {{( - 1)}^l}\sqrt {{{\left( {{\varepsilon _A} - {\varepsilon _B}} \right)}^2} + 4{w_m}(t_{i},\vec{\xi_{i}},\vec{k})}   } \right] 
\label{eq:full-shrun-eigen} 
\end{eqnarray}
The corresponding orthogonal  eigenvectors are
\begin{eqnarray}
\left| {{\phi}(E_{sh,n}(\vec k))} \right\rangle  = C_{A_3}(E_{sh;n}(\vec k))\left[ \begin{array}{*{20}{l}}
\left(\frac{C_{A_{1}}(E_{sh;n}(\vec k))}{C_{A_{3}}(E_{sh;n}(\vec k))}~
\frac{C_{A_{2}}(E_{sh;n}(\vec k))}{C_{A_{3}}(E_{sh;n}(\vec k))}~
1~\right)~
C \frac{\eta^{*}(E_{sh;n}(\vec k)) }{\left| \eta(E_{sh;n}(\vec k)) \right|}\left(\frac{C^{*}_{A_{1}}(E_{sh;n}(\vec k))}{C^{*}_{A_{3}}(E_{sh;n}(\vec k))}~
 \frac{C^{*}_{A_{2}}(E_{sh;n}(\vec k))}{C^{*}_{A_{3}}(E_{sh;n}(\vec k))}~
1\right)\end{array} \right]^T
\label{eq:shrunk-eigenstates-0} 
\end{eqnarray}
where $C_{A_3}(E_{sh,n}(\vec k))$ can be found from orthogonality condition. Also,
\begin{eqnarray}
{C_{{A_1}}}(E_{sh,i}(\vec k)) = \frac{{ - \left( {{G_{22}} - {w_i}(\vec k)} \right){G_{13}} + {G_{12}}{G_{23}}}}{{\left( {{G_{11}} - {w_i}(\vec k)} \right)\left( {{G_{22}} - {w_i}(\vec k)} \right) - {{\left| {{G_{12}}} \right|}^2}}}{C_{{A_3}}}(E_{sh,i}(\vec k))\nonumber\\
{C_{{A_2}}}(E_{sh,i}(\vec k)) = \frac{{ - \left( {{G_{11}} - {w_i}(\vec k)} \right){G_{23}} + {G_{21}}{G_{13}}}}{{\left( {{G_{11}} - {w_i}(\vec k)} \right)\left( {{G_{22}} - {w_i}(\vec k)} \right) - {{\left| {{G_{12}}} \right|}^2}}}{C_{{A_3}}}(E_{sh,i}(\vec k)).
\label{eq:CA1-CA2-coef} 
\end{eqnarray}
To find ${ \xi }_{B}(E^{0}_{n}(\vec k))$ it has been used symmetry condition in Eq.~\ref{eq:shrunk-eigenvalue-sepr}
\begin{eqnarray}
{C_{{B_m}}}(E_{sh;i}(\vec k)) =C{e^{i\varphi (E_{sh;i}(\vec k))}}C_{{A_m}}^*(E_{sh;i}(\vec k)).
\label{eq:CB1-CB2-coef} 
\end{eqnarray}
Replacing Eq.~\ref{eq:CB1-CB2-coef}  into second equation of Eq.~\ref{eq:couple-eigen} we get
\begin{eqnarray}
{e^{i\varphi (E_{sh;i}(\vec k))}}& =&  \frac{{{\eta ^*}(E_{sh;i}(\vec k))}}{{\left| {\eta (E_{sh;i}(\vec k))} \right|}}\frac{{{C_{{A_3}}}(E_{sh;i}(\vec k))}}{{C_{{A_3}}^*(E_{sh;i}(\vec k))}} \quad ; \quad C = \frac{{E^{0}_{sh,i}(\vec k) - {\varepsilon _1}(\vec k)}}{{E^{0}_{sh,i}(\vec k) - {\varepsilon _2}(\vec k)}}\nonumber\\
\eta (E_{sh;i}(\vec k)) &= & - {t_1}\left( {{d_3}\frac{{C_{{A_1}}^*(E_{sh;i}(\vec k))}}{{C_{{A_3}}^*(E_{sh;i}(\vec k))}} + {d_2}\frac{{C_{{A_2}}^*(E_{sh;i}(\vec k))}}{{C_{{A_3}}^*(E_{sh;i}(\vec k))}} + {\tau _2}} \right).
\label{eq:eta-def} 
\end{eqnarray}
It is easy to show that 
$\left| {\eta (E_{sh;i}(\vec k))} \right| = {w_i}(\vec k) $
\section{Coupled Li-C$_6$ dispersion relations }  
 By applying the following unitary transformation,
$P_{0N}^\dag {{\hat H}_N}{P_{0N}}\left( {P_{0N}^\dag \left| {{\Psi _{\vec k,n}}(\vec r)} \right\rangle } \right) = {E_n}(\vec k)P_{0N}^\dag \left| {{\Psi _{\vec k,n}}(\vec r)} \right\rangle $,
  where ${\hat P_{0N}}$ is the operator that diagonalize Eq.~\ref{eq:non-perturb-sch}, 
the Schr\"odinger Eq.~\ref{eq:k-schr-1} is written in a new matrix representation  as 
\begin{eqnarray}
\left( {\begin{array}{*{20}{c}}
E_{Li,0}(\vec k) &\vline& \gamma _{1}(\vec k)&{{\gamma _2}(\vec k)}&{{\gamma _3}(\vec k)}&{{\gamma _4}(\vec k)}&{{\gamma _5}(\vec k)}&{{\gamma _6}(\vec k)}\\
\hline
{\gamma _1^ * (\vec k)}&\vline& E_{sh;1}(\vec k) &0&0&0&0&0\\
{\gamma _2^ * (\vec k)}&\vline& 0&E_{sh;2}(\vec k)  &0&0&0&0\\
{\gamma _3^ * (\vec k)}&\vline& 0&0&E_{sh;3}(\vec k)  &0&0&0\\
{\gamma _4^ * (\vec k)}&\vline& 0&0&0&  E_{sh;4}(\vec k)  &0&0\\
{\gamma _5^ * (\vec k)}&\vline& 0&0&0&0&  E_{sh;5}(\vec k)  &0\\
{\gamma _6^ * (\vec k)}&\vline& 0&0&0&0&0&  E_{sh;6}(\vec k) 
\end{array}} \right)\left( \begin{array}{*{20}{c}}
A_{0}({E_i}(\vec k))\\
A_{1}({E_i}(\vec k))\\
A_{2}({E_i}(\vec k))\\
A_{3}({E_i}(\vec k))\\
A_{4}({E_i}(\vec k))\\
A_{5}({E_i}(\vec k))\\
A_{6}({E_i}(\vec k))
\end{array} \right) = E_{i}(\vec k)\left( \begin{array}{*{20}{c}}
A_{0}(E_{i}(\vec k))\\
A_{1}(E_{i}(\vec k))\\
A_{2}(E_{i}(\vec k))\\
A_{3}(E{_i}(\vec k))\\
A_{4}(E_{i}(\vec k))\\
A_{5}(E_{i}(\vec k))\\
A_{6}(E_{i}(\vec k))
\end{array} \right) 
\label{eq:schro-inbasisofH0N}
\end{eqnarray}
where the relation between the column matrix eigenstate of Eq.~\ref{eq:schro-inbasisofH0N}, 
$A(E_{i}(\vec k))$, and  the eigenstates of Eq.~\ref{eq:k-schr-1}, $\mathscr{C}$ is 
\begin{equation}
A={P}^{\dagger}_{0N}\mathscr{C},\;\;\mbox{and}\;\;\;
{{\rm A}_j}(E_i(\vec k)) = \frac{{\gamma _j^*}}{{{E_i}(\vec k) - E_{sh,j}(\vec k)}}A_{0}(E_i(\vec k))
\label{eq:trans-eigenstates}
\end{equation} 
in which $A_{0}(E_i(\vec k))$ is determined from the normalization condition, and also
\begin{equation}
{\gamma _i}(\vec k) =-t_{1}^{LiC} \sum\limits_{m = 1}^3 (C_{A_m}({E_{sh,i}}(\vec k)) e^{i\vec k.{{\vec \zeta }_{A_m}}}+C_{B_m}({E_{sh,i}(\vec k)}) e^{i\vec k.{{\vec \zeta }_{B_m}}})
\label{eq:H1-new-components}
\end{equation}
where $\vec{\zeta}_{A_m}$ is a vector that connects $Li$ to the $A_m$ carbon atom and $\vec{\zeta
}_{B_m}$ is a vector which connects $Li$ to the $B_m$ carbon atom.

 At the $\Gamma$ point, $\gamma_{6}(0)=-\sqrt{6}\; t_{1}^{LiC}$ and $\gamma_{i}(0)=0$ for $i=1,...,5$. These results show that just the
isolated intercalant band, $E_{Li,0}(0)$ and the lowest valance band, $E_{sh,6}(0)$, are mutually affected. The energies of these bands are, with $E_0(0)\equiv E_{+}$, $E_6(0)\equiv E_{-}$, 
\begin{eqnarray}
{E_{\pm}(0)} = \frac{1}{2}\left( E_{Li,0}(0) + E_{sh,6}(0) \right) \pm 
   \sqrt {{{\frac{1}{4}\left[ E_{Li,0}(0) -E_{sh,6}(0)  \right]}^2} + 6(t_{1}^{LiC})^2}  
\end{eqnarray}
and other shrunk graphene bands Eq.~\ref{eq:gamma-energy-shrunk} remain unchanged. This means 
the energy gap at $\Gamma$, $E_g(0)$, depends only on the nearest neighbor hopping difference 
rather than on $t_{1}^{LiC}$.
 That is because the overlap between the Li $s$ band and the valance band  of uncoupled shrunken graphene (which is linear combination of $s$ and $f$, $\left| {{\phi_{6}}(0)} \right\rangle$) 
is significant while others are zero.

For general ${\vec k}$ vectors it is challenging to obtain an exact analytical expression 
for the full Hamiltonian in Eq.~\ref{eq:k-schr-1} and it would not be transparent anyway. 
One can use perturbation theory to obtain useful results.
 For $H_{ND}=H_{0D}+H_{1D}$ we can use non degenerate perturbation theory, obtaining
\begin{eqnarray}
H_{ND}(\vec k)\left| {{\psi _N}(E_{i})} \right\rangle  = E_{i}(\vec k)\left| {{\psi _N}(E_{i})} \right\rangle 
\label{eq:HND}
\end{eqnarray}
where expansion of $E_{i}(\vec k)$ in terms of perturbation parameter is
$
E_{i}(\vec k) = E_i^0(\vec k) + \mathcal{E}_i^1(\vec k) + \mathcal{E}_i^2(\vec k)  +  ... .
$
From non degenerate perturbation we have
\begin{eqnarray}
\mathcal{E}_i^1(\vec k) = \left\langle \phi_{i}\left| {{H_1D}} \right|{\phi_{i}} \right\rangle,\;\;\;
\mathcal{E}_i^2(\vec k) = \sum_{j \ne i} \frac{{\left| {\left\langle {{\phi_{i}}\left| {{H_1D}} \right|{\phi_{j}}} \right\rangle } \right|}^2}{E_i^0(\vec k) - E_j^0(\vec k)} 
\label{eq:HND-2}
\end{eqnarray}
 where $\left|{\phi_{i}} \right\rangle$ is $i$th eigenstate of diagonal $H_{0D}$. 
 Perturbed system eigenstates up to first order are
\begin{eqnarray}
|\psi_{N}(E_{i}(\vec k))>=\left|\phi_{i}(E^{0}_{i}(\vec k)\right\rangle+\sum_{j\ne i} \frac{1}{E_i^0 - E_j^0}\left| \phi^{0}_{j}(E_{j}(\vec k)\right\rangle.
\label{eq:1-2-6-on}
\end{eqnarray}
 Non degenerate perturbation theory can be used in Eq.~\ref{eq:schro-inbasisofH0N} for completely filled or empty bands that  are far from lithium band, $E_{Li,0}(\vec k)$, and also without overlap. Therefore, except $E_{sh,2}$ and $E_{sh,3}$ which are nearly degenerate with lithium band in some regions, non degenerate approximation can be used for other bands of $H_{0N}$. We denote Hamiltonian in Eq.~\ref{eq:schro-inbasisofH0N} as $H_{DN}$. This Hamiltonian can be separated to $H_{ND}=H_{0D}+H_{1D}$ where
\begin{eqnarray}
H_{0D} = \left( \begin{array}{*{20}{c}}
E_{Li,0}(\vec{k})&\vline& 0&{{\gamma _2}(\vec k)}&{{\gamma _3}(\vec k)}&0&0&0\\
\hline
0&\vline&E_{sh,1}(\vec{k})&0&0&0&0&0\\
\gamma _2^ * (\vec k)&\vline& 0&E_{sh,2}(\vec{k})&0&0&0&0\\
\gamma _3^ * (\vec k)&\vline& 0&0&E_{sh,3}(\vec{k})&0&0&0\\
0&\vline& 0&0&0&E_{sh,4}(\vec{k})&0&0\\
0&\vline& 0&0&0&0&E_{sh,5}(\vec{k})&0\\
0&\vline& 0&0&0&0&0&E_{sh,6}(\vec{k})
\end{array} \right).
\end{eqnarray}
\begin{eqnarray}
H_{1D} = \left( \begin{array}{*{20}{c}}
0&\vline& {{\gamma _1}(\vec k)}&0&0&{{\gamma _4}(\vec k)}&{{\gamma _5}(k)}&{{\gamma _6}(\vec k)}\\
\hline
\gamma _1^ * (\vec k)&\vline& 0&0&0&0&0&0\\
0&\vline& 0&0&0&0&0&0\\
0&\vline& 0&0&0&0&0&0\\
\gamma _4^ * (\vec k)&\vline& 0&0&0&0&0&0\\
\gamma _5^ * (\vec k)&\vline& 0&0&0&0&0&0\\
\gamma _6^ * (\vec k)&\vline& 0&0&0&0&0&0
\end{array} \right).
\label{eq:gamma-def}
\end{eqnarray} 
introducing below coefficients 
\begin{eqnarray}
{c_2} &=&E_{Li,0}(\vec k)+E_{sh,2}(\vec k)+E_{sh,3}(\vec k)\nonumber\\
{c_1} &=& -(E_{Li,0}(\vec k)E_{sh,2}(\vec k)+E_{Li,0}(\vec k)E_{sh,3}(\vec k)+E_{sh,2}(\vec k)E_{sh,3}(\vec k)-|\gamma_2(\vec k)|^{2}-|\gamma_3(\vec k)|^{2})\nonumber\\
{c_0} &=&  E_{Li,0}(\vec k)E_{sh,2}(\vec k)E_{sh,3}(\vec k)-E_{sh,3}(\vec k)|\gamma_2(\vec k)|^{2}
-E_{sh,2}(\vec k)|\gamma_3(\vec k)|^{2}\nonumber\\
\Pi  &=& (q + i\sqrt {{p^3} - {q^2}} ),\;\;
q = \frac{{{c_0}}}{2} + \frac{{{c_1}{c_2}}}{6} + \frac{{c_2^3}}{{27}},\;\;\;
p = \frac{{{c_1}}}{3} +\frac{{c_2^2}}{9}
\end{eqnarray}
Non trivial  eigenstate  of $H_{0D}$ are 
\begin{eqnarray}
E_{0}^{0}(\vec k)=  \frac{{{c_2}}}{2} + \Pi  + {\Pi ^*},\;\;
 E_{2}^{0}(\vec k)=  \frac{{{c_2}}}{2} + {e^{i2\pi /3}}\Pi  + {e^{ - i2\pi /3}}{\Pi ^*},\;\;
 E_{3}^{0}(\vec k)=   \frac{{{c_2}}}{2} + {e^{ - i2\pi /3}}\Pi  + {e^{i2\pi /3}}{\Pi ^*}
\label{eq:equ-1-3}
\end{eqnarray}
and corresponding eigenstates are
\begin{eqnarray}
\left| {{\phi_{0}}(E^{0}_{0}(\vec k))} \right\rangle  = (C_{p1}(E^{0}_{0}(\vec k))\; 0\;C_{p2}(E^{0}_{0}(\vec k))C_{p3}(E^{0}_{0}(\vec k))0\;\;0\;\;0)^{T}\nonumber\\
\left| {{\phi_{2}}(E^{0}_{2}(\vec k))} \right\rangle  = (C_{p1}(E^{0}_{2}(\vec k))\; 0\;C_{p2}(E^{0}_{2}(\vec k))C_{p3}(E^{0}_{2}(\vec k))0\;\;0\;\;0)^{T}\nonumber\\
\left| {{\phi_{3}}(E^{0}_{3}(\vec k))} \right\rangle  = (C_{p1}(E^{0}_{3}(\vec k))\; 0\;C_{p2}(E^{0}_{3}(\vec k))C_{p3}(E^{0}_{3}(\vec k))0\;\;0\;\;0)^{T}\nonumber\\
\end{eqnarray}
It is easy to show that ${\mathcal E}_1^0=0$. So up to second order perturbation parameter the eigenenergies are
\begin{eqnarray}
{E_{0}}(\vec k)& =&{E_{0}^{0}}(\vec k) -|C_{p1}(E_{0}^{0})|^{2}\left(\frac{{{{\left| {{\gamma _1}} \right|}^2}}}{{E_{sh,1} - E_{0}^{0}}}+\frac{{{{\left| {{\gamma _4}} \right|}^2}}}{{E_{sh,4} - E_{0}^{0}}}+\frac{{{{\left| {{\gamma _5}} \right|}^2}}}{{E_{sh,5} - E_{0}^{0}}}+\frac{{{{\left| {{\gamma _6}} \right|}^2}}}{{E_{sh,6} - E_{0}^{0}}}\right)\nonumber\\
{E_{1}}(\vec k) &=&{E_{sh,1}}(\vec k) +|C_{p1}(E_{0}^{0})|^{2}{\frac{{{{\left| {{\gamma _1}} \right|}^2}}}{{E_{sh,1} - E_{0}^{0}}}} +|C_{p1}(E_{2}^{0})|^{2}{\frac{{{{\left| {{\gamma _1}} \right|}^2}}}{{E_{sh,1} - E_{2}^{0}}}} +|C_{p1}(E_{3}^{0})|^{2}{\frac{{{{\left| {{\gamma _1}} \right|}^2}}}{{E_{sh,1} - E_{3}^{0}}}} \nonumber\\
{E_{2}}(\vec k) &=&{E_{2}^{0}}(\vec k) -|C_{p1}(E_{2}^{0})|^{2}\left(\frac{{{{\left| {{\gamma _1}} \right|}^2}}}{{E_{sh,1} - E_{2}^{0}}}+\frac{{{{\left| {{\gamma _4}} \right|}^2}}}{{E_{sh,4} - E_{2}^{0}}}+\frac{{{{\left| {{\gamma _5}} \right|}^2}}}{{E_{sh,5} - E_{2}^{0}}}+\frac{{{{\left| {{\gamma _6}} \right|}^2}}}{{E_{sh,6} - E_{2}^{0}}}\right)\nonumber\\
{E_{3}}(\vec k) &=&{E_{3}^{0}}(\vec k) -|C_{p1}(E_{3}^{0})|^{2}\left(\frac{{{{\left| {{\gamma _1}} \right|}^2}}}{{E_{sh,1} - E_{3}^{0}}}+\frac{{{{\left| {{\gamma _4}} \right|}^2}}}{{E_{sh,4} - E_{3}^{0}}}+\frac{{{{\left| {{\gamma _5}} \right|}^2}}}{{E_{sh,5} - E_{3}^{0}}}+\frac{{{{\left| {{\gamma _6}} \right|}^2}}}{{E_{sh,6} - E_{3}^{0}}}\right)\nonumber\\
{E_{4}}(\vec k) &=&{E_{sh,4}}(\vec k) +|C_{p1}(E_{0}^{0})|^{2}{\frac{{{{\left| {{\gamma _4}} \right|}^2}}}{{E_{sh,4} - E_{0}^{0}}}} +|C_{p1}(E_{2}^{0})|^{2}{\frac{{{{\left| {{\gamma _4}} \right|}^2}}}{{E_{sh,4} - E_{2}^{0}}}} +|C_{p1}(E_{3}^{0})|^{2}{\frac{{{{\left| {{\gamma _4}} \right|}^2}}}{{E_{sh,4} - E_{3}^{0}}}} \nonumber\\
{E_{5}}(\vec k) &=&{E_{sh,5}}(\vec k) +|C_{p1}(E_{0}^{0})|^{2}{\frac{{{{\left| {{\gamma _5}} \right|}^2}}}{{E_{sh,5} - E_{0}^{0}}}} +|C_{p1}(E_{2}^{0})|^{2}{\frac{{{{\left| {{\gamma _5}} \right|}^2}}}{{E_{sh,5} - E_{2}^{0}}}} +|C_{p1}(E_{3}^{0})|^{2}{\frac{{{{\left| {{\gamma _5}} \right|}^2}}}{{E_{sh,5} - E_{3}^{0}}}} \nonumber\\
{E_{6}}(\vec k) &=&{E_{sh,6}}(\vec k) +|C_{p1}(E_{0}^{0})|^{2}{\frac{{{{\left| {{\gamma _6}} \right|}^2}}}{{E_{sh,6} - E_{0}^{0}}}} +|C_{p1}(E_{2}^{0})|^{2}{\frac{{{{\left| {{\gamma _6}} \right|}^2}}}{{E_{sh,6} - E_{2}^{0}}}} +|C_{p1}(E_{3}^{0})|^{2}{\frac{{{{\left| {{\gamma _6}} \right|}^2}}}{{E_{sh,6} - E_{3}^{0}}}} \nonumber\\
\label{eq:perturbed-band-energies}
\end{eqnarray}
\section{ Bogoliubov-de Gennes Transformation }
The electron-electron interaction part of Hamiltonian, $H_{P}$, 
in the mean field approximation becomes
\begin{eqnarray}
 \hat H_P^{MF} =    \frac{1}{2}\sum\limits_{i\alpha \sigma } {\sum\limits_{j\beta \sigma '} \Delta _{i\alpha j\beta }^{\sigma \sigma '}\hat c_{i\alpha \sigma }^\dag \hat c_{j\beta \sigma '}^\dag  + h.c. + {F_0}} 
=   \frac{1}{2}  \sum_{\vec k\alpha \sigma \beta \sigma '} \Delta _{\alpha \beta }^{\sigma \sigma '}(\vec k)\hat c_{\alpha \sigma }^\dag(\vec k) \hat c_{\beta \sigma '}^\dag (-\vec k) + h.c. + {F_0} 
\label{eq:meanfield-Hp}
\end{eqnarray}
in which   $\Delta _{i\alpha j\beta }^{\sigma \sigma '} 
  = U_{i\alpha ,j\beta }^{\sigma \sigma '} \left\langle {{{\hat c}_{i\alpha \sigma }}{{\hat c}_{j\beta \sigma '}}} \right\rangle$ is the matrix of order parameters  in real space.
   Fourier transformation of real space order parameters are given by
\begin{eqnarray}
 \Delta _{\alpha \beta }^{\sigma \sigma '}(\vec k) 
  = \frac{1}{ N}\sum_{ij} \Delta _{i\alpha j\beta }^{\sigma \sigma '} 
    e^{i\vec k. ({\vec r}_{i\alpha}-{\vec r}_{j\beta})}.  
\label{eq:site-orderparameter-def}
\end{eqnarray}
here Latin subscripts, $\alpha$ and $\beta$ refers to $A_i$ or $B_i$ subsites.
 The interacting Hamiltonian in Nambu space is
 \begin{eqnarray}
 {{\hat H}_{SU}} = \sum_{\vec k} {{{\hat \Psi }^\dag }(\vec k)} {H_{SU}}(\vec k)\hat \Psi (\vec k)
\label{eq:Hsu-operator}
\end{eqnarray}
where 
$ \hat \Psi^{\dagger} (\vec k)=(c^{\dagger}_{0\uparrow}(\vec k)\;c^{\dagger}_{1\uparrow}(\vec k)~,...,~c^{\dagger}_{6\uparrow}(\vec k)\;
c_{0\downarrow}(-\vec k)\;c_{1\downarrow}(-\vec k)\;,...,
c_{6\downarrow}(-\vec k))$  where $Li$, $A_1$, $A_2$, $A_3$, $B_1$, $B_2$ and $B_3$ are labeled by $0,~1,2,3,~4,5,6$ respectively, and
\begin{eqnarray}
H_{SU}(\vec k) = \left( \begin{array}{*{20}{c}}
H_{N}(\vec k)&H_{P}(\vec k)\\
H^{\dag }_{P}(\vec k)&- H^{*}_N( -\vec k)
\end{array} \right).
\label{eq:Hsu-matrix}
\end{eqnarray}
The coupling is given by
\begin{eqnarray}
{H_P}(\vec k) = \left( {\begin{array}{*{20}{c}}
0&\vline&0&0&0&\vline& 0&0&0\\
\hline
0&\vline& 0&0&0&\vline& \Delta^{\uparrow\downarrow}_{A_1B_1}(\vec k)&\Delta^{\uparrow\downarrow}_{A_1B_2}(\vec k)&\Delta^{\uparrow\downarrow}_{A_1B_3}(\vec k)\\
0&\vline& 0&0&0&\vline& \Delta^{\uparrow\downarrow}_{A_2B_1}(\vec k)&\Delta^{\uparrow\downarrow}_{A_2B_2}(\vec k)&\Delta^{\uparrow\downarrow}_{A_2B_3}(\vec k)\\
0&\vline& 0&0&0&\vline& \Delta^{\uparrow\downarrow}_{A_3B_1}(\vec k)&\Delta^{\uparrow\downarrow}_{A_3B_2}(\vec k)&\Delta^{\uparrow\downarrow}_{A_3B_3}(\vec k)\\
 \hline
0&\vline& \Delta^{\uparrow \downarrow*}_{A_1B_1}(\vec k)&\Delta^{\uparrow \downarrow *}_{A_2B_1}(\vec k)&\Delta^{\uparrow \downarrow *}_{A_3B_1}(\vec k)&\vline& 0&0&0\\
0&\vline& \Delta^{\uparrow \downarrow *}_{A_1B_2}(\vec k)&\Delta^{\uparrow \downarrow *}_{A_2B_2}(\vec k)&\Delta^{\uparrow\downarrow *}_{A_3B_2}(\vec k)&\vline& 0&0&0\\
0&\vline& \Delta^{\uparrow\downarrow *}_{A_1B_3}(\vec k)&\Delta^{\uparrow\downarrow *}_{A_2B_3}(\vec k)&\Delta^{\uparrow\downarrow *}_{A_3B_3}(\vec k)&\vline& 0&0&0
\end{array}} \right)  
\label{eq:pairing-sitespace-Hp}
\end{eqnarray}
for singlet $\Delta^{\uparrow\downarrow}_{\beta\alpha}(\vec k)=\Delta^{\uparrow\downarrow}_{\alpha\beta}(-\vec k)=-\Delta^{\downarrow\uparrow}_{\alpha\beta}(-\vec k)={\Delta^{\uparrow\downarrow}_{\alpha\beta}}^{*}(\vec k)$. The order parameters
 according to (main text) Fig.~4 are
\begin{eqnarray}
\Delta^{\uparrow\downarrow}_{A_1B_1}(\vec k)&=&
\Delta''_{1} e^{i{\vec k}.{\vec \tau}_{1}}, ~~~
\Delta^{\uparrow\downarrow}_{A_1B_2}(\vec k)=\Delta_{2} e^{i{\vec k}.{\vec\delta}_{2}},~~~
\Delta^{\uparrow\downarrow}_{A_1B_3}(\vec k)=\Delta^{'}_{3} e^{i{\vec k}.{\vec \delta}_{3}};\nonumber\\
\Delta^{\uparrow\downarrow}_{A_2B_1}(\vec k)&=&\Delta^{'}_{2} e^{i{\vec k}.{\vec \delta}_{2}},~~~
\Delta^{\uparrow\downarrow}_{A_2B_2}(\vec k)=\Delta^{''}_{3} e^{i{\vec k}.{\vec \tau}_{3}},~~~
\Delta^{\uparrow\downarrow}_{A_2B_3}(\vec k)=\Delta_{1} e^{i{\vec k}.{\vec \delta}_{1}};\nonumber\\
\Delta^{\uparrow\downarrow}_{A_3B_1}(\vec k)&=&
\Delta_{3} e^{i{\vec k}.{\vec \delta}_{3}},~~~
\Delta^{\uparrow\downarrow}_{A_3B_2}(\vec k)=
\Delta^{'}_{1} e^{i{\vec k}.{\vec \delta}_{1}},~~~
\Delta^{\uparrow\downarrow}_{A_3B_3}(\vec k)=\Delta^{''}_{2} e^{i{\vec k}.{\vec \tau}_{2}}. 
\label{eq:singlet-cond}
\end{eqnarray}
$U_{<iA_1\uparrow j B_1\downarrow>}=U_{<iA_2\uparrow j B_2\downarrow>}=U_{<iA_3\uparrow j B_3\downarrow>}g_{1}$
Quasiparticle energies are obtained by unitary transformation in the seven band space 
\begin{eqnarray}
{\hat H}_{SU} = \sum_{\vec k} {\hat \Psi }^{\dag }(\vec k) Q\left[ {{Q^\dag }{H_{SU}}(\vec k)Q} \right]{Q^\dag }\hat \Psi (\vec k)
 = \sum_{\vec k} \Lambda ^{\dag }(\vec k)H_{SU}^N(\vec k)\Lambda (\vec k) 
\label{eq:trans-Hsu}
\end{eqnarray}
where in matrix notation,
\begin{eqnarray}
H_{SU}^N(\vec k)= \left( {\begin{array}{*{20}{c}}
{{H_{ND}}(k)}&{{H_{PD}}(\vec k)}\\
{H_{PD}^\dag (\vec k)}&{ - H_{ND}^*( -\vec k)}
\end{array}} \right)
,~~~Q = \left( \begin{array}{*{20}{c}}
{\hat P}_{N}(\vec k)&{\mathord{\buildrel{\lower3pt\hbox{$\scriptscriptstyle\frown$}} 
\over 0} }\\
{\mathord{\buildrel{\lower3pt\hbox{$\scriptscriptstyle\frown$}} 
\over 0} }&{\hat P_N^ * ( -\vec k)}
\end{array} \right).
\label{eq:Hsu-bands-space}
\end{eqnarray}
 ${\hat P}_{N}(\vec k)$ is a $7\times7$ matrix where each column is one of 
the perturbed normal state eigenvectors of Eq. \ref{eq:k-schr-1}, thus with 
matrix elements given by 
$[{\hat P}^{\dag}_{N}]_{i,j}  = \mathscr{C}_{i}({E_j(\vec k)})$. 
$H_{ND}(\vec k)$ is the corresponding diagonal seven band Hamiltonian.
Here $\hat P_N^*( -\vec k) = {{\hat P}_N}(\vec k)$. In the normal band space the matrix 
 elements of the off-diagonal array are given by
\begin{eqnarray}
[{H_{DP}}(\vec k)]_{i,j} =\Delta _{ij}(\vec k)
= \sum_{\alpha  = 1}^9 \Omega _{ij}^\alpha (\vec k)\Delta^{\alpha } 
\label{eq:band-orderparameter-matrix}
\end{eqnarray}
Using the fact that the gap is small, 
applying perturbation up to second order in the order parameter gives quasiparticle energies
\begin{eqnarray}
E_{m,s}^Q (\vec k)= s\left( E_{m}(\vec k)   + \sum_{i = 1}^7 \frac{{\left| \Delta_{mi}(\vec k) \right|}^2}{{E_m}(\vec k) + E_{i}(\vec k) }  \right)  \qquad  s =  \pm 1
\label{eq:quasi-su-spectrum}
\end{eqnarray}
where $s=1$ is for particles and $s=-1$ for holes.
\section{Superconducting States}
By minimizing the quasiparticle free energy with respect to nearest neighbor order 
parameters the gap equation is obtained.  The free energy is
\begin{eqnarray}
F =  - \frac{2 }{\beta }\sum_{\vec k} \sum_{n = 1}^7 {\ln \left[ {2\cosh (\frac{ E_n^Q}{ 2 k_{B}T})} \right]  + {F_0}} .   
\label{eq:quasi-su-free}
\end{eqnarray}
 where $F_0$ is the system condensation energy, 
\begin{eqnarray}
 {F_0} =  - \frac{1}{2}\sum_{i\alpha \sigma } \sum_{j\beta \sigma '} \Delta _{i\alpha j\beta }^{\sigma \sigma '}{G^{\sigma \sigma '}_{i\alpha j\beta }}^{\dagger}
   =  -2N\sum_{\alpha  = 1}^9 J_{\alpha }(\Delta^{\alpha})^{2} 
\label{eq:min-free-energy}
\end{eqnarray}
where $J_{1}=J_{2}=J_{3}=\frac{1}{ g_{1}}$ and $J_{4}=J_{5}=J_{6}=J_{7}=J_{8}=J_{9}=\frac{1}{ g_{0}}$.
The linearized gap equation, obtained by minimizing free energy of the system, is
\begin{equation}
J_{\beta }\Delta^{\beta } =- \frac{1}{2N}\sum_{\alpha =1}^9 \left[ \sum_{\vec k} 
   \sum_{n = 1}^7 \sum_{i = 1}^7 \frac{\tanh (\frac{{ E_n^Q}}{ 2 k_{B}T})}{E_{n}(\vec k) 
  + E_{i}(\vec k) } \left( \Omega _{ni}^\alpha (\vec k)\Omega _{ni}^{ * \beta }(\vec k) 
  + \Omega _{ni}^\beta (\vec k)\Omega _{ni}^{ * \alpha }(\vec k) \right) \right] \Delta^{\alpha }
 \equiv - \sum_{\alpha  = 1}^9 \Gamma _{\beta \alpha } \Delta^{\alpha }.
\label{eq:gap-Equation}
\end{equation}
We have used that at $T_{c}$, where $|\Delta_{ij}|^2$ can be neglected, ${ E_n^Q}\rightarrow
E_n$. 

In the general case, ( see main text Fig.~ 4)
, the 
system is invariant under interchange  
$2\rightleftharpoons 5$, 
$3\rightleftharpoons 6$ and 
$4 \rightleftharpoons 7$, 
which means  $\Delta^{'}_{i}\rightleftharpoons \Delta_{i}$. 
These relations correspond to 
$\Delta^{4}\rightleftharpoons \Delta^{7}$, 
$\Delta^{5}\rightleftharpoons \Delta^{8}$ and 
$\Delta^{6}\rightleftharpoons \Delta^{9}$ in Eq.~\ref{eq:gap-Equation}. 
This symmetry and that of the symmetric $\Gamma$ matrix
 $\Gamma_{\beta\alpha}=\Gamma_{\alpha\beta}$ 
 allows   Eq.~\ref{eq:gap-Equation} to be written in matrix form as   
\begin{eqnarray}
\left[ \begin{array}{*{20}{c}}
A_{3 \times 3}&B_{3 \times 3}&B_{3 \times 3}\\
B_{3 \times 3}&C_{3 \times 3}&D_{3 \times 3}\\
B_{3 \times 3}&D_{3 \times 3}&C_{3 \times 3}
\end{array} \right]\left( \begin{array}{*{20}{c}}
g_{1}V_{1}\\
g_{0}V_{2}\\
g_{0}V_{3}
\end{array} \right) =- \left( \begin{array}{*{20}{c}}
V_{1}\\
V_{2}\\
V_{3}
\end{array} \right)
\label{eq:matrix-form-gap-eq}
\end{eqnarray}
where
\begin{eqnarray}
A_{3 \times 3} = \left[ \begin{array}{*{20}{c}}
\Gamma _{11}&\Gamma _{12}&\Gamma _{12}\\
\Gamma _{12}&\Gamma _{11}&\Gamma _{12}\\
\Gamma _{12}&\Gamma _{12}&\Gamma _{11}
\end{array} \right],\;    C_{3 \times 3} = \left[ \begin{array}{*{20}{c}}
\Gamma _{44}&\Gamma _{45}&\Gamma _{45}\\
\Gamma _{45}&\Gamma _{44}&\Gamma _{45}\\
\Gamma _{45}&\Gamma _{45}&\Gamma _{44}
\end{array} \right],\;\;
B_{3 \times 3} = \left[ \begin{array}{*{20}{c}}
\Gamma_{14}&\Gamma_{15}&\Gamma_{15}\\
\Gamma_{15}&\Gamma_{14}&\Gamma_{15}\\
\Gamma_{15}&\Gamma_{15}&\Gamma_{14}
\end{array} \right],\;\;    D_{3 \times 3} = \left[ \begin{array}{*{20}{c}}
\Gamma_{47}&\Gamma_{48}&\Gamma_{48}\\
\Gamma_{48}&\Gamma_{47}&\Gamma_{48}\\
\Gamma_{48}&\Gamma_{48}&\Gamma_{47}
\end{array} \right]
\label{eq:matrix-gap-elements}
 \end{eqnarray}
Eq.~\ref{eq:matrix-form-gap-eq} can be written as a non-Hermitian eigenvalue problem
\begin{eqnarray}
\left[ \begin{array}{*{20}{c}}
{\kappa A}&{\kappa B}&{\kappa B}\\
B&C&D\\
B&D&C
\end{array} \right]\left( \begin{array}{*{20}{c}}
{{g_1}{V_1}}\\
{{g_0}{V_2}}\\
{{g_0}{V_3}}
\end{array} \right) =  - \frac{1}{g_{0}}\left( \begin{array}{*{20}{c}}
g_{1}V_{1}\\
g_{0}V_{2}\\
g_{0}V_{3}
\end{array} \right)
\label{eq:gapequation-matrix-unhermit-form}
\end{eqnarray}
where $\kappa =\frac{g_{1}}{g_{0}}$. 
 Substates $V_1$, $V_2$ and $V_3$  in  gap equation Eq. \ref{eq:gapequation-matrix-unhermit-form}, 
cannot be have different symmetries, so each of the eigenvectors of 
Eq.~\ref{eq:gapequation-matrix-unhermit-form} can be expressed
in the compact form 
\begin{eqnarray}
[\Phi_{n}]^{T}=[\alpha_{sy}V_{sy}\quad \beta_{sy}V_{sy}\quad \gamma_{sy}V_{sy}]
\label{eq:gap-general-function}
\end{eqnarray}
where subscript $sy$ refers to one of the $s$-wave, $d_{xy}$-wave or $d_{x^2-y^2}$-wave symmetries, 
and the coefficients $\alpha_{sy}$, $\beta_{sy}$ and $\gamma_{sy}$ are to be determined. 
By inserting eigenvectors from Eq.~\ref{eq:gap-general-function} into the gap equation 
Eq.~\ref{eq:gapequation-matrix-unhermit-form} one finds
\begin{eqnarray}
\frac{\alpha_{sy}}{\beta_{sy}}=\frac{\alpha_{sy}}{\gamma_{sy}}\;,\qquad\;\;
\frac{\gamma_{sy}}{\beta_{sy}}=\frac{\beta_{sy}}{\gamma_{sy}}\;, \qquad
J_{sy}=c_{sy}+\frac{\beta_{sy}}{\gamma_{sy}}d_{sy}+\frac{\alpha_{sy}}{\gamma_{sy}}b_{sy}
\label{eq:coef-general-function}
\end{eqnarray}
where $c_{sy}=c_{s}, c_d $, $b_{sy}=b_{s}, b_d $, $d_{sy}=d_{s}, d_d $ and 
$J_{sy}=-\frac{1}{g_0}$ for each symmetry. Eq.~\ref{eq:coef-general-function} has two classes of solutions 
\begin{eqnarray}
\beta_{sy}&=&-\gamma_{sy}\equiv 1\quad \Rightarrow\quad \alpha_{sy}=0,~~J_{sy}^{0}=c_{sy}-d_{sy},    \nonumber \\
\beta_{sy}&=&+\gamma_{sy}\equiv 1\quad \Rightarrow\quad 
   b_{sy}\alpha_{sy}^2+(c_{sy}+d_{sy}-\kappa a_{sy})\alpha_{sy}-2\kappa b_{sy}=0
\label{eq:coef-secondorder}
\end{eqnarray}
In the limiting case 
of pristine graphene, 
the quadratic equation in Eq.~\ref{eq:coef-secondorder}, has two temperature independent solutions 
$\alpha_{sy}=1$ and  $\alpha_{sy}=-2$. The last solution in addition to island 
states $\alpha_{sy}=0$ are in fact, orthogonal states where they are linear combination of 
the aforementioned $\Phi_{0n}$ and $\Phi_{1n}$. But in the general case of  symmetry breaking
LiC$_{6}$  characterized by gap equation Eq.~\ref{eq:matrix-form-gap-eq},  
the quadratic equation in Eq.~\ref{eq:coef-secondorder},
has two temperature dependent solutions,
\begin{eqnarray} 
\alpha _{sy}^{\pm} =  \frac{J_{sy}^{\pm} - c_{sy} - d_{sy}}{b_{sy}},~~
J_{sy}^{\pm} =\frac{1}{2}\left({\kappa {a_{sy}} + {c_{sy}} + {d_{sy}} 
             \pm \sqrt {8\kappa b_{sy}^2 + {{\left[ {{c_{sy}} + {d_{sy}} 
             - \kappa {a_{sy}}} \right]}^2}} } \right)
\label{eq:quad-eigenvalue}
\end{eqnarray}

\begin{table}[ht]
\centering
\begin{tabular}{|l|l|l l|l l| l l|l l|l |} 
\bottomrule [1pt]
\hline
$n$ & $0$ & $1$ & $2$ & $3$ & $4$ & $5$ & $6$ & $7$ & $8$ & $9$\\  
\hline
$t_{0n}^{CC}$ & $\epsilon_{c}=-0.77$ & $t_1=2.93$ & $t'_1=0.94t_1$ & $t_2=-0.22$ & $t'_2=0.94t_2 $ & $t_3=0.28$&$t'_3\approx t_3$ & $t_4=-0.03$ & $ t'_4\approx t_4$ & $t_5=-0.05$\\ 
\hline
\bottomrule [1pt]
\hline
$m$ & $0$ & $1$ & $2$ & $3$ & $4$ & &&&& \\ 
\hline
$t_{0m}^{LiLi}$ &$\epsilon_{Li}=1.1$&$-0.30$&$0.09$& $0.04$ &$-0.03 $&& & &&\\ 
\hline
$t_{0m}^{LiC}$ &$-$&$0.30$&&  & &&&  && \\ 
\hline
\end{tabular}
\caption{The C-C hopping parameters (eV) for LiC$_6$ are denoted by $t^{CC}_{0n}$ where 
the index $n$ indicates $n$-th neighbours.  In the Li plane, Li-Li hopping parameters 
are denoted by $t_{0m}^{LiLi}$ where $m$ is $m$-th Li neighbor of central Li. 
The Li-C hopping parameter is  $t_{0m}^{LiC}$.} 
\label{table:shrunk-hoppping} 
\end{table}


\begin{thebibliography}{0}
\bibitem{Wang2012}
Wang, Q. Y.  {\it et al.}
  Interface-Induced High-Temperature Superconductivity in Single Unit-Cell FeSe Films on
   SrTiO$_3$. Chin. Phys. Lett. {\bf 29}, 037402 (2012).

\bibitem{He2013}
He, S. L.  {\it et al.}
  Phase diagram and electronic indication of high-temperature superconductivity at 65 K 
  in single-layer FeSe films. Nat. Mater. {\bf 12}, 605 (2013).

\bibitem{Cao2018}
 Cao, Y. {\it et al.}
  Unconventional superconductivity in magic-angle graphene superlattices.
Nature  {\bf 556},  43-50 (2018).
\bibitem{Weller-naturephysics2005}
 Weller,  T. E. {\it et al.} Superconductivity in the 
intercalated graphite compounds C$_6$Yb and C$_6$Ca.
 Nature Phys. {\bf 1}, 39 (2005).
 
 \bibitem{Ludbrook2015}
  Ludbrook,  B.M. {\it et al. }
  Evidence for superconductivity in Li-decorated monolayer graphene. 
 Proc. Natl. Acad. Sci. USA {\bf 112}, 11795-11799 (2015).

\bibitem{Palinkas2017}
Palinkas, A. {\it et al. }
 Novel graphene/Sn and graphene/SnO$_x$ hybrid nanostructures: induced superconductivity
  and band gaps revealed by scanning probe measurements.
 Carbon {\bf 124}, 611 (2017).
 
\bibitem{Tiwari2017}
 Tiwari, A. P. {\it et al. }
 Superconductivity at 7.4K in few layer graphene by Li intercalation.
 J. Phys.: Condens. Matt. {\bf 29}, 445701 (2017).
 
\bibitem{Woo2017}
 Woo, S. {\it et al. }
 Temperature-dependent transport properties of graphene deorated by alkali metal
  adatoms (Li,K).
 Appl. Phys. Lett. {\bf 111}, 263502 (2017).


\bibitem{Uchoa}
 Uchoa, B. \& Castro Neto, A. 
   Superconducting States of Pure and Doped Graphene. 
 Phy. Rev. Lett. {\bf 98}, 146801 (2007).  

\bibitem{Nandkishore}
 Nandkishore, R., Levitov, L. S. \& Chubukov, A. V. 
  Chiral superconductivity from repulsive interactions in doped graphene 
 Nat. Phys. {\bf 8}, 158–163 (2012).                
                                                          
\bibitem{Nandkishore2014}
 Nandkishore, R., Thomale R. \& Chubukov A. V. Superconductivity from weak repulsions in hexagonal lattice systems. Phys. Rev. B 89, 144501 (2014).  

 \bibitem{Blackschaffer2007} 
 Black-Schaffer, A. M. \& Doniach, S. 
   Resonating valence bonds and mean field d-wave superconductivity in graphene. 
 Phys. Rev. B {\bf 75}, 134512 (2007).                                                     

\bibitem{Kiesel}
 Kiesel M. L. , Platt C., Hanke W. , Abanin D. A., \& Thomale R. 
  Competing many-body instabilities and unconventional superconductivity in graphene. 
 Phys. Rev. B {\bf 86}, 020507R (2012).  

\bibitem{Ma}
 Ma, T., Yang, F., Yao, H. \& Lin H. Q. 
  Possible triplet $p+ip$ superconductivity in graphene at low filling. 
 Phys. Rev. B {\bf 90}, 245114 (2014). 
 
\bibitem{Profeta}
 Profeta, G., Calandra, M. \& Mauri, F. Phonon-mediated  superconductivity in graphene by Lithium deposition. Nat. Phys. {\bf 8}, 131-134 (2012).

\bibitem{Wong2017} 
 Wong, C. H., Lortz, R.,  Buntov, E. A.,  Kasimova, R. E. \&
 Zatsepin,  A. F. 
 A theoretical quest for high temperature superconductivity on the example of
  low-dimensional carbon structures.
 Sci. Rep. {\bf 7}, 15805 (2017).
 
\bibitem{Hou}
 Hou C. Y., Chamon, C., \& Mudry C. 
  Electron Fractionalization in Two-Dimensional Graphenelike Structures. 
 Phys. Rev. Lett. {\bf 98}, 186809 (2007).

\bibitem{Long-Hua}
 Wu, L. -H. \& Hu, X.
  Topological Properties of Electrons in Honeycomb Lattice with Detuned Hopping Energy.
  Sci. Rep. {\bf 6}, 24347 (2016). 

\bibitem{Guzman}
Guzman, D. M., Alyahyaei H. M. \&  Jishi, R. A.
  Superconductivity in graphene-lithium.
 2D Materials {\bf 1}, (2014) 021005.

\bibitem{Wallace1947}
 Wallace, P. R. The Band Theory of Graphite. 
Phys. Rev. {\bf 71}, 622 (1947).
\bibitem{Reich2002}

 Reich, S.,  Maultzsch, J., Thomsen,  C. \&  Ordej\'on, P.
 Tight-binding description of graphene.
  Phys. Rev. B {\bf 66}, 035412 (2002).
  
\bibitem{Kundu2011}
 Kundu, R. Tight Binding Parameters for Graphene. 
Mod. Phys. Lett. B {\bf 25}, 163 (2011).

\bibitem{Jung2013}
 Jung J. \& MacDonald,  A. H.
  Tight-binding model for graphene $\pi$-bands from maximally localized Wannier functions.
 Phys.Rev. B {\bf 87}, 195450 (2013).
 
\bibitem{Zheng2016}
 Zheng, J. -J. \&  Margine, E. R. First-principles calculations of the superconducting properties in Li-decorated monolayer graphene within the anisotropic Migdal-Eliashberg formalism.  Phys.Rev. B {\bf 94}, 064509 (2016).
\end{thebibliography}
\end{document}